\documentclass[prb,aps,twocolumn,amsmath,amssymb,floatfix,superscriptaddress]{revtex4}

\usepackage{graphicx}
\usepackage{xcolor}
\usepackage{float}
\usepackage{braket}
\usepackage{color}
\definecolor{dred}{rgb}{0.75,0,0}
\usepackage{soul}
\usepackage[colorlinks=true, citecolor=blue, urlcolor=blue]{hyperref}
\begin{document}

\title{Correlated quasiperiodicity enables efficient thermoelectric energy conversion}

\author{Swaraj Biswas}

\email{swarajbiswas972@gmail.com}

\affiliation{Physics and Applied Mathematics Unit, Indian Statistical Institute, 203 Barrackpore Trunk Road, Kolkata-700 108, India} 

\author{Santanu K. Maiti}

\email{santanu.maiti@isical.ac.in}

\affiliation{Physics and Applied Mathematics Unit, Indian Statistical Institute, 203 Barrackpore Trunk Road, Kolkata-700 108, India} 


\begin{abstract}

We investigate a route to enhanced thermoelectric energy conversion in nanoscale systems by exploiting a correlated quasiperiodic 
energy landscape in a one-dimensional chain coupled to source and drain reservoirs. The considered modulation generates a highly 
non-uniform electronic transmission spectrum, providing favorable conditions for achieving a large thermoelectric figure of merit. 
By systematically tuning the incommensurability parameter, a variety of quasiperiodic configurations are explored, several of which 
yield high values of the figure of merit exceeding $2$. Electronic transport properties are evaluated within a tight-binding framework 
using the non-equilibrium Green's function formalism, while the thermoelectric coefficients, including electrical conductance, Seebeck
coefficient, and electronic thermal conductance, are determined through the Landauer approach. The underlying quasiperiodic potential 
belongs to the Aubry-Andr\'{e}-Harper (AAH) family and exhibits a weakly varying spatial profile, leading to transmission characteristics 
that are favorable for thermoelectric optimization. The influence of phonon thermal conductance on the overall energy-conversion 
efficiency is also analyzed in detail. For the sake of completeness, we also critically inspect the effect of conductor
to electrode coupling and the coupling asymmetry on $ZT$. We also check the thermoelectric response of conventional AAH system and an 
elaborate comparison is made with our chosen quantum system. Our findings highlight the potential of correlated quasiperiodic 
nanostructures as promising candidates for efficient thermoelectric applications.

\end{abstract}

\maketitle

\section{Introduction}

Thermoelectricity is a phenomenon in which heat is converted into electricity\cite{1,2,3}. For the past few decades, thermoelectricity is an active field of research. It may provide an effective way to enhance energy efficiency and promote environmental sustainability\cite{4}, particularly in the context of waste heat recovery\cite{5}. Thermoelectric materials can offer an eco-friendly solution to get rid of 
modern-day industrial challenges by providing a green, non-mechanical approach for generating electrical power and maintaining thermal balance. 
Thermoelectric (TE) device performance is characterized by a factor called the figure of merit\cite{6,7,8,9,noza} defined as $ZT = G S^{2}T/k$ = $GS^{2}T/(k_e + k_{ph})$, where $G$ and $S$ are electrical conductance and Seebeck coefficient, respectively, $k$ denotes total thermal conductance, which is written as a sum of two parts $k_e$ and $k_{ph}$. Here, $k_e$ and $k_{ph}$ denote electronic and phonon thermal conductances, respectively, and $T$ is the equilibrium temperature of the system. High values of $ZT$ ($ \geq $ 2) are required for efficient energy conversion. Although theoretical models\cite{10,11,12} outline valuable insights to enhance $ZT$, practical issues like material fabrication, stability, and durability concerns remain barriers. However, to align the mismatch between experimental outcomes\cite{13,14,15} 
of $ZT$ ($<$ 1) with theoretical predictions remain a key focus in ongoing progress in thermoelectric engineering and development in material designing.

Possible ways to enhance the value of $ZT$ are: (i) increasing electrical conductance and Seebeck coefficient, and (ii) decreasing total thermal conductance. A high value of $ZT$ indicates high electrical conduction and low thermal conduction. However, it is difficult to satisfy these conditions simultaneously in conventional bulk materials. From the Wiedemann-Franz law\cite{17,18}, it is well known that in bulk systems $G$ and $k$ cannot be changed independently.  Therefore, in bulk systems $ZT$ values are much smaller and hence they are not efficient thermoelectric materials\cite{19}. 

Low-dimensional systems consist of a few number of atoms, hence less scattering probability, which consequences high values of $G$. Similarly, $k$ is expected to be low due to a small number of atoms. Low-dimensional materials also offer high values of $S$. As $ZT$ is directly proportional to $S^2$, we need to have a high value of $S$ as well for an efficient energy conversion\cite{20,21}. Variation of electronic transmission function {\Large$\tau$}($E$) ($E$ represents the energy) above and below $E_F$ within a small energy window $K_BT$ plays a crucial part in enhancing the Seebeck coefficient, thereby enhancing $ZT$. The above criteria is satisfied in low-dimensional materials because of quantum confinement effects that result in significant modification of their electronic density of states (DOS). Due to above benefits low-dimensional materials are expected to have large values of $ZT$ compared to bulk \cite{22,23,24,25,26,27,28,29,30}.

Although many ideas are put forward using different types of simple and complicated systems, further investigation is still required. Specifically, correlated disordered systems draw attention, where atomic potentials are interrelated. In recent decades, quasiperiodic 
crystals emerge as a major research topic\cite{31,32,33,34,35,skm1,skm2}. Quasiperiodic crystals are defined as a special class of correlated disordered solid materials that exhibit long-range ordering with aperiodic atomic arrangement. Unlike ordinary crystals, quasiperiodic structures do not possess translational symmetry but still the pattern is deterministic, not random. It is intermediate between fully periodic and completely random disordered lattices\cite{anderson,ran1}. 

Quasiperiodic systems may offer possible routes for efficient thermoelectric applications because of their distinctive electronic 
and phononic properties~\cite{th1}. Unlike periodic systems, quasiperiodic lattices exhibit a rich energy spectrum consisting of multiple 
sub-bands and gaps\cite{31,32,33,34,35,skm1,skm2}. Such spectral features can produce a strong asymmetry in electronic transmission 
profile, leading to an enhanced thermopower when the Fermi energy lies near a sub-band edge. At the same time, the electrical conductance 
can remain sufficiently high. The other important aspect of quasiperiodic structures is their ability to reduce phonon transport. Due to 
the absence of translational symmetry, phonon scattering is enhanced, which can reduce thermal conductance~\cite{phcnd1}. This is one 
of the key requirements to achieve a favorable thermoelectric response.

There are different kinds of quasiperiodic systems like Fibonacci chain, Aubry-André-Harper (AAH) lattices, weakly varying AAH chain 
etc\cite{43,44,45,46,47}. Among these, the weakly varying AAH potential is a special type of incommensurate quasiperiodic disorder, 
where the modulation potential changes gradually over the lattice sites\cite{48,49,slv1,slv2}. Different configurations are generated 
by tuning the incommensurate factor, which determines the smoothness of spatial variation of the potential. In this case, 
for any finite value of the disorder strength or incommensurate factor, there always exists a mixed phase, i.e., both 
localized and extended states occur simultaneously. So, `weakly varying AAH potential' can induce energy dependent localization instead 
of complete localization over the entire spectrum. This causes asymmetry in transmission spectrum, which is one of the primary requirements 
for getting significantly large $ZT$. This feature is not observed in `random' (uncorrelated) disordered or conventional one-dimensional 
1D AAH lattices with nearest-neighbor hopping, where all eigenstates become localized beyond a critical disorder strength~\cite{32,34,skm1}. 
In the former case, viz., the weakly varying potential systems, Fermi energy can be tuned to sit near the sub-band edges to harvest the 
extended carriers along with steep density of state variations. This model can be realized experimentally, and attracts significant 
research interest.

Recent advancements in low-dimensional thermoelectrics have established that nano-waveguides and thermoelectric 
metamaterials can engineer the electron and phonon transport through structural modulation to increase thermoelectric 
efficiency~\cite{nw1,nw2}. In such systems, spatial geometry modulation can induce scattering and interference of electron and phonon 
waves, and affects their transport characteristics. Unlike the conventional geometrical modulation, here in the presence work, the spatial
correlations in site energies in the form of weakly varying AAH model can generate energy-dependent localization relevant to thermoelectric
transport phenomena.

The primary objectives of our study are: (i) to obtain highly asymmetric transmission lineshape in the presence of disorder, (ii) to 
observe the behavior of all the thermoelectric quantities, and (iii) to achieve high values of $ZT$ by imposing the correlated disorder, 
including both electronic and phononic contributions.

The remaining portion is arranged as follows. In Sec. II, we discuss the theoretical formalism to calculate the thermoelectric quantities. 
Numerical results are elaborately analyzed in Sec. III. Experimental realization of weakly varying potential is briefly discussed in 
Sec. IV, and finally in Sec. V, we summarize our essential findings and provide an outlook for future studies.  

\section{Quantum system and theoretical method}

\subsection{Thermoelectric junction setup and tight-binding Hamiltonian}

In our setup, we consider a finite one-dimensional nanoscopic weakly varying AAH chain consisting of $N$ number of lattice sites,
\begin{figure}[htbp]
\includegraphics[width=8cm]{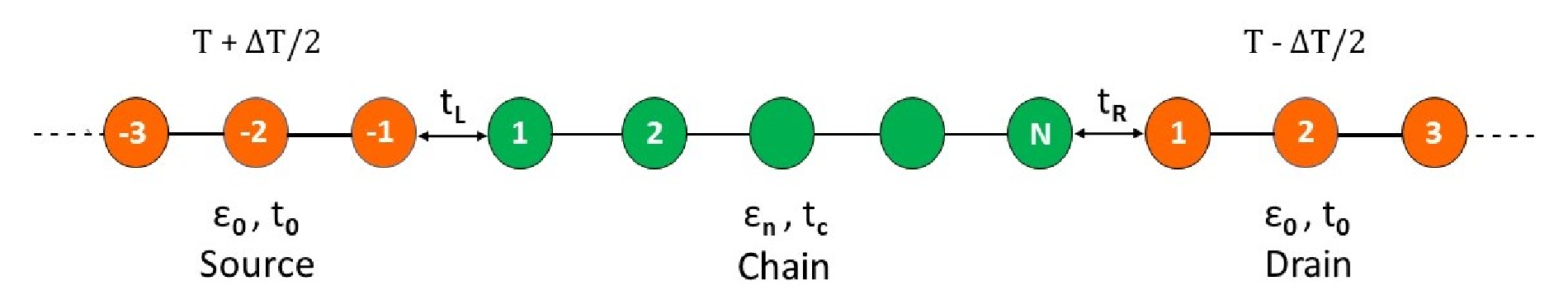}
\caption{\small(Color online). Schematic representation of a 1D chain coupled to two 1D heat baths, source and drain. The heat baths are 
kept at temperatures $T+\Delta T/2$ and $T-\Delta T/2$, respectively. $T$ is the equilibrium temperature of the system, and $\Delta T$ is
the temperature difference.}
\label{fig.1}
\end{figure} 
embedded between two 1D semi-infinite electrodes, namely, the source and the drain as shown in Fig.~\ref{fig.1}. Electrodes are assumed 
to be identical and kept at temperatures $T + \Delta T/2$ and $T - \Delta T/2$ respectively. Here we assume $\Delta T$ to be so small 
to reside in the linear response regime.

We describe our model quantum system in a tight-binding (TB) framework. The total Hamiltonian for the system can be written as
\begin{equation}
H=H_C+H_S+H_D+H_T,\end{equation} where $H_C$, $H_S$, $H_D$, and $H_T$ are different sub-Hamiltonians that correspond to the chain, source, drain, and coupling, respectively. We will describe each part one by one here. 

\begin{figure}[htbp]
	\includegraphics[width=7.5cm]{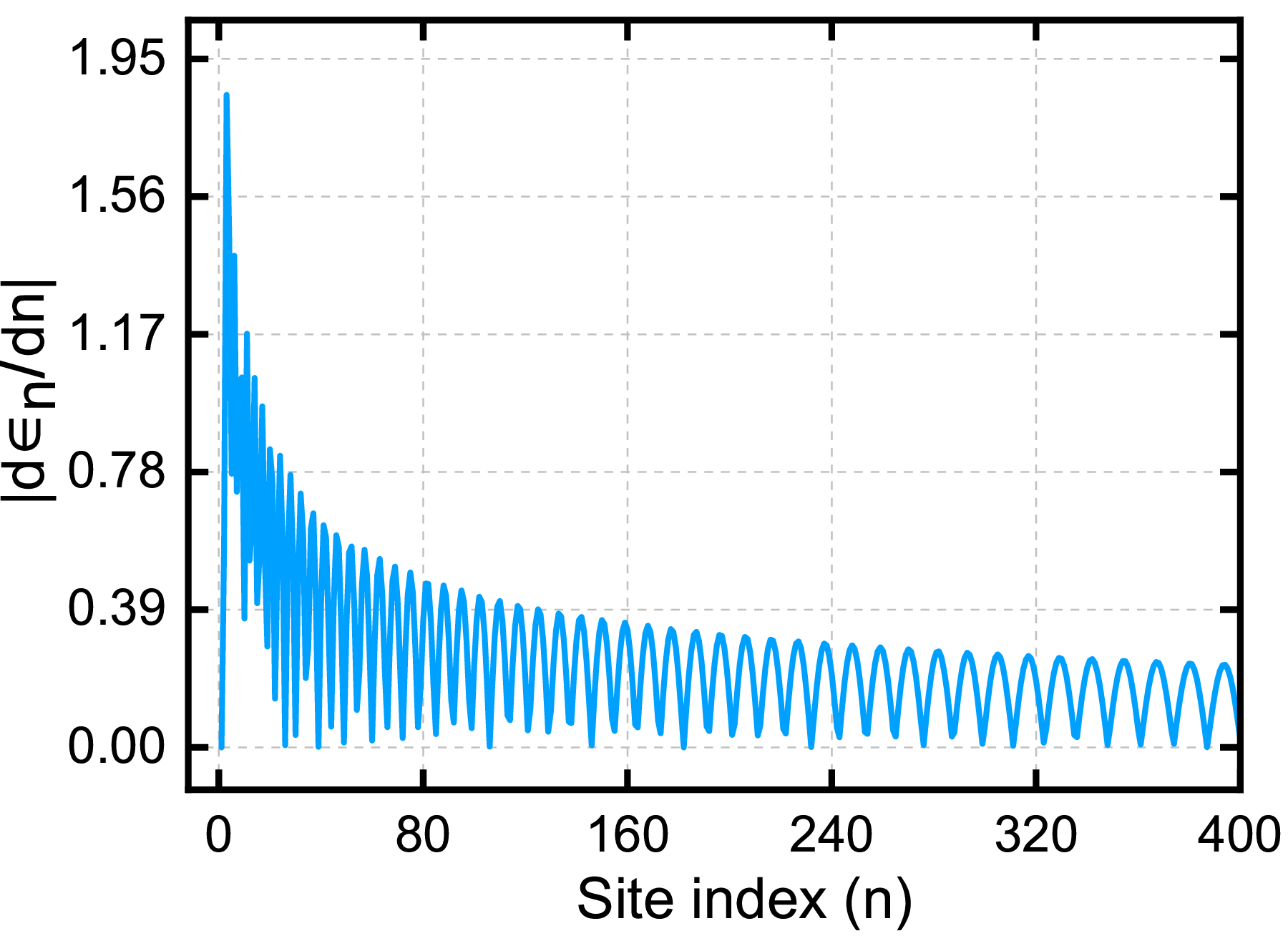}
	\caption{\small(Color online). Absolute derivative of site potential as a function of site index.}
	\label{fig.2}
\end{figure}

The Hamiltonian $H_C$ for the chain is written as
\begin{equation}
H_C = \sum_{n=1}^N\epsilon_nc_n^\dagger c_n+t_c\sum_{n=1}^{N-1}(c_{n+1}^\dagger c_n+h.c.).
\end{equation}
Here, $c_n^\dagger$ and $c_n$ are electronic creation and annihilation operators at the $n$th site of the chain, respectively, and
$t_c$ is the nearest-neighbor hopping (NNH) amplitude in the chain. On-site energies are modulated as weakly varying AAH 
potential~\cite{48,49,slv1,slv2} which is described as
\begin{equation}
\epsilon_n=W\cos (2\pi n^\nu),
\end{equation}
where $W$ is the disorder amplitude, and $\nu$ be the incommensurate factor that varies from $0$ to $1$. 
The chain is disordered for any other values of $\nu$ except $0$ and $1$. For $0$ and $1$, the chain becomes perfect. The term 
``weakly varying" is used in the sense that potential varies weakly and incommensurately with the system size~\cite{48,49,slv1,slv2}. 
This behavior is shown in Fig.~\ref{fig.2}, where it can be seen that the derivative of the site potential almost approaches zero for 
large systems. Here we want to add that, for our chosen system the site energies are explicitly modulated in the above 
mentioned form, and this is not the confinement-induced modulation in site potentials.

The Hamiltonians $H_S$ and $H_D$, for the source and drain, respectively, are written as
\begin{equation}
    H_S= \sum_{l\leq-1}\epsilon_0a_l^\dagger a_l+t_0\sum_{l\leq-1}(a_{l+1}^\dagger a_l+h.c.),
\end{equation} 
and 
\begin{equation}
H_D=\sum_{l\geq1}\epsilon_0b_l^\dagger b_l+t_0\sum_{l\geq1}(b_{l+1}^\dagger b_l+h.c.),
\end{equation}
where $\epsilon_0$ and $t_0$ are the site energies and hopping strengths, respectively, of the source and drain. 
$a_l^\dagger$ and $a_l$ are electronic creation and annihilation operators for the source, respectively. $b_l$ is used in a same way like $a_l$ operator, for the drain.

The coupling Hamiltonian $H_T$ reads as \begin{equation}
    H_T=(t_L c_1^\dagger a_{-1}+t_Rb_1^\dagger c_N+h.c.).
\end{equation}
Here, $t_L$ is the coupling strength between the chain and the source, and $t_R$ denotes the coupling strength between the chain and the drain.

\subsection{Theoretical prescription}

\subsubsection{Calculation of electronic transmission probability}

The expression of the electronic transmission probability {\Large$\tau$}($E$), evaluated using the non-equilibrium Green's function technique (NEGF)\cite{51,59}, is given as \begin{equation}
    \scalebox{1.4}{$\tau$}(E) = Tr\left[\Gamma_L G^{r} \Gamma_R G^{a}\right],
\end{equation}
where, $\Gamma_L$ and $\Gamma_R$ denote the coupling matrices and $G^r$ is the retarded Green's function defined as 
\begin{equation}
 G^{r} = \left[E\mathbb{I}-H_c - \Sigma_L-\Sigma_R\right]^{-1}.
\end{equation}
Here, $E$ is the incoming electron's energy and $\mathbb{I}$ is the ($N\times N$) identity matrix. $\Sigma_L$ and $\Sigma_R$ denote self-energy matrices, corresponding to source and drain, respectively, which can be evaluated using the term $\Sigma_{L/R}^\prime$ = $\frac{t_{L/R}^2}{t_0}$ exp$\left[-i\cos^{-1}\left(\frac{(E-\epsilon_0)}{2t_0}\right)\right]$. $G^{a}$ is the conjugate transpose of $G^{r}$. $\Gamma_L$ and $\Gamma_R$ can be computed from the expression $\Gamma_{L/R}$ = i$\left[\Sigma_{L/R}- \Sigma_{L/R}^\dagger\right]$. All the matrices have the 
dimensions of ($N\times N$).

\subsubsection{Determination of ZT}

Efficiency of a thermoelectric material is determined by $ZT = GS^2T/k$, where meaning of the terms $G$, $S$, and $k$ are already written before. Phononic contribution to the thermal conductance becomes too small\cite{22,53} in the nanoscale regime, due to small system size. However, for the accurate estimation of $ZT$, the effect of phonon thermal conductance $k_{ph}$, should be taken into account. In our study, we calculate $ZT$, including $k_{ph}$, to examine its effect on $ZT$. Theoretical method for calculating $k_{ph}$ will be discussed in the 
next part. 

Thermoelectric quantities, apart from $k_{ph}$, are evaluated using the Landauer prescription \cite{22,52,54} given below as:
\begin{equation}\label{eq:10}
    G=2e^2 L_0 ,
\end{equation}
\vspace{-20pt}
\begin{equation}\label{eq:11}
    S=\frac{L_1}{eTL_0},
\end{equation}
\vspace{-15pt}
\begin{equation}
    k_e=\frac{2}{T}(L_2-\frac{L_1^2}{L_0}),
\end{equation}
where $L_0$, $L_1$, and $L_2$ are Landauer integrals \cite{22,52,54} defined as
\begin{equation}\label{eq:13}
    L_n=-\frac{1}{h}\int\scalebox{1.4}{$\tau$}(E)\left(\frac{\partial f}{\partial E}\right)(E-E_F)^n dE.
\end{equation}
Here, {\Large$\tau$}($E$) represents two-terminal electronic transmission probability, $f$ is the equilibrium Fermi-Dirac function, $E_F$ is the Fermi energy, $e$ is the electronic charge, and $h$ is the Planck's constant.
\subsubsection{Estimation of phonon thermal conductance}
Residing in the linear response regime, the expression of phonon thermal conductance ($k_{ph}$) within the NEGF formalism can be 
written as\cite{55,56} 
\begin{equation}
k_{ph} = \frac{\hbar}{2\pi} \int_{0}^{\omega_{c}}\scalebox{1.4}{$\tau_{\scriptscriptstyle ph}$}(\omega)
\left(\frac{\partial f_{BE}}{\partial T}\right)\omega d\omega.
\end{equation}
Here, $\omega$ and $\omega_{c}$ denote phonon frequency and phonon cutoff frequency, respectively; $f_{BE}$ represents Bose-Einstein distribution function, $\hbar$ is the reduced Planck constant, and {\Large$\tau_{\scriptscriptstyle ph}$}($\omega$) is the phonon transmission probability across the chain.\\
\textbullet\ \textbf{Determination of {\Large$\tau_{\scriptscriptstyle ph}$}}: Phonon transmission probability {\Large$\tau_{\scriptscriptstyle ph}$} can be calculated using the NEGF formalism as 
\begin{equation}
    \scalebox{1.4}{$\tau_{\scriptscriptstyle ph}$} = Tr\left[\Gamma_S G_{ph} \Gamma_D G_{ph}^\dagger\right],
\end{equation}
where, $\Gamma_{S}$ and $\Gamma_{D}$ are thermal broadening matrices, and $G_{ph}$ represents the Green's function for the phonon in the chain, expressed as 
\begin{equation}
    G_{ph} = \left[\mathbb{M}\omega^2-\mathbb{K} - \Sigma_S-\Sigma_D\right]^{-1}.
\end{equation}
Here, $\mathbb{M}$ is a diagonal matrix, where each element $\mathbb{M}_{nn}$ denotes mass of the $n$th atom in the chain and $\mathbb{K}$ is the spring constant matrix, where diagonal element $\mathbb{K}_{nn}$ is proportional to the restoring force acting on the $n$th atom due to $(n-1)$th and $(n+1)$th atoms and off-diagonal element $\mathbb{K}_{ni}$ denotes spring constant between $n$th and $i$th neighbors. $\Sigma_S$ and $\Sigma_D$ denote self-energy matrices for the source and drain, respectively. Expression for $\Gamma_{S}$ and $\Gamma_{D}$ can be written as $\Gamma_{S/D}$ = i$\left[\Sigma_{S/D}-\Sigma_{S/D}^\dagger\right]$. Self-energy matrices is computed using the term $\Sigma_{S/D}^\prime$ = $-K_{S/D}$ exp$\left[2i \sin^{-1}(\frac{\omega}{\omega_c})\right]$, where $K_{S/D}$ denotes spring constant at the source/drain and chain interface. Self-energy matrices $\Sigma_S$ and $\Sigma_D$, $\mathbb{M}$, and $\mathbb{K}$ have same dimensions.

We evaluate the spring constants for the 1D heat baths and the 1D chain, using the Harrison's interatomic potential\cite{56,57}. In the absence of any transverse interaction for a 1D system\cite{58}, the expression of the spring constant for the source, drain and chain can be written as $K$ = 3$d$$c_{11}$/{16}, where $d$ is the interatomic spacing and $c_{11}$ is the elastic constant of the corresponding material. The cutoff frequency is determined using $\omega_c$ = 2$\sqrt{K/M}$, with $M$ being the mass.

\section{Numerical results and discussion}

Here, we will elaborately discuss the results one by one. Before starting, it is important to mention the physical parameters taken in 
this numerical analysis and these are kept constant throughout the work. Set of parameters considered as: $\epsilon_0=0\,$eV, 
$t_0=3\,$eV, $W = 1\,$eV, $t_c = 1\,$eV, $t_L=t_R=0.75\,$eV. A $40$-site chain is taken and we fix the equilibrium temperature at 
$200\,$K. Other variable parameters, those are not constant, are mentioned in subsequent parts.

The central idea to increase energy efficiency is to obtain a `highly asymmetric electronic transmission spectrum'. In case of a 
weakly varying AAH potential, both localized and extended states appear simultaneously for any finite value of incommensurate factor 
$\nu$, which causes asymmetry in the electronic transmission function. So, before starting our discussion about the thermoelectric 
quantities and $ZT$, it is necessary to analyze the electronic transmission profile. We consider different configurations of this 
potential by tuning the incommensurate factor. We elaborately discuss the effect of phonon thermal conductance on $ZT$ in the final 
part of our analysis. 

First of all, we discuss the electronic transmission function {\Large$\tau$}($E$) for two different values of $\nu$.
\begin{figure}[htbp]
\includegraphics[width=7cm]{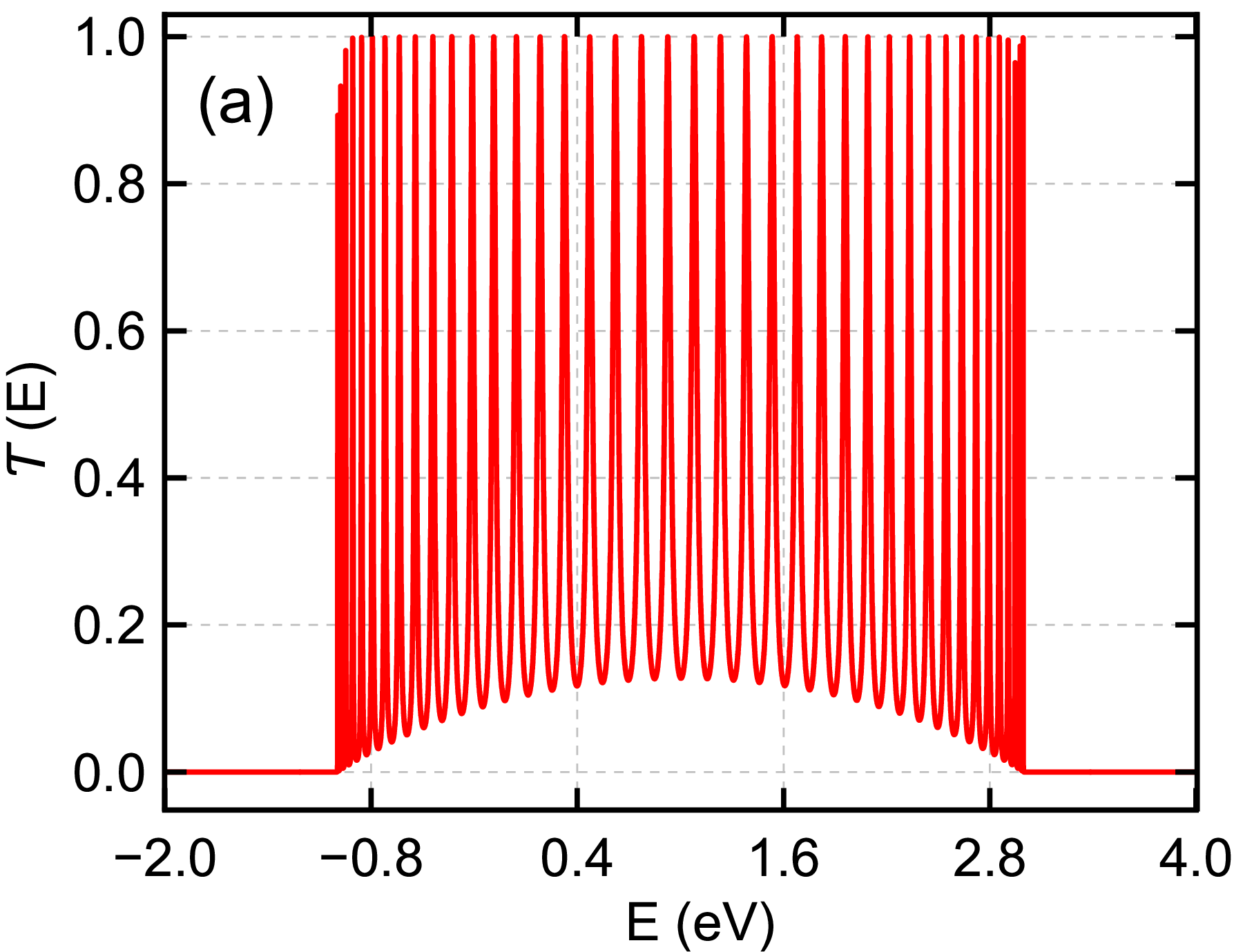}\vskip\baselineskip 
\includegraphics[width=7cm]{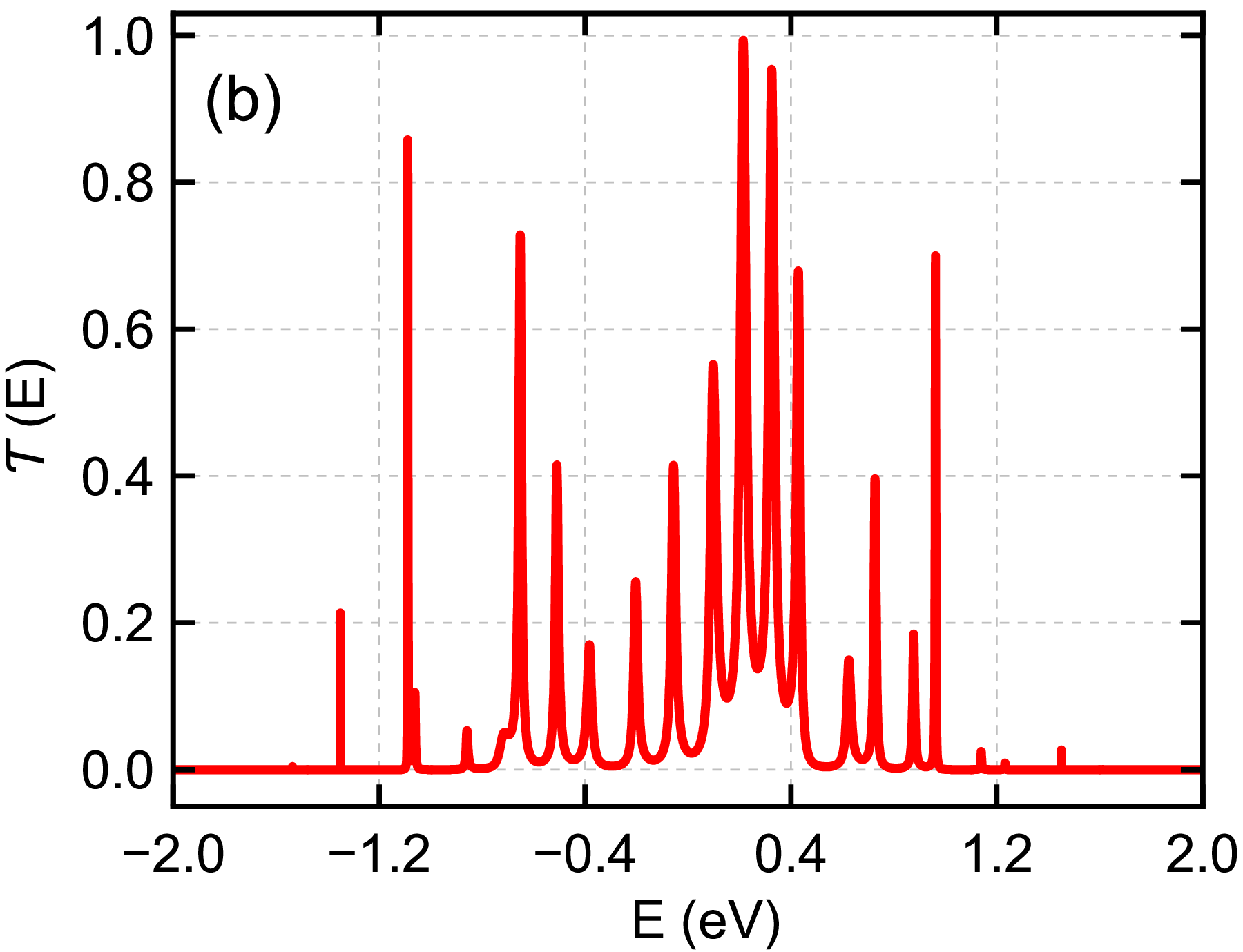}
\caption{\small(Color online). Electronic transmission probability as a function of energy for (a) $\nu=0\,$ and (b) $\nu=0.55\,$ 
with $N=40\,$ and $W=1\,$eV.} 
\label{fig.3}
\end{figure}
Transmission profile is uniform throughout the entire energy window for the perfect case, i.e., $\nu=0\,$, as shown in Fig.~\ref{fig.3}(a). In a perfect lattice, the absence of spatial variation of onsite potential allows electrons to propagate freely without any backscattering. So, all energy eigenchannels are equally probable to transmit electrons and we get identical peaks for all energies.
\begin{figure*}[htbp]
    \includegraphics[width=16cm]{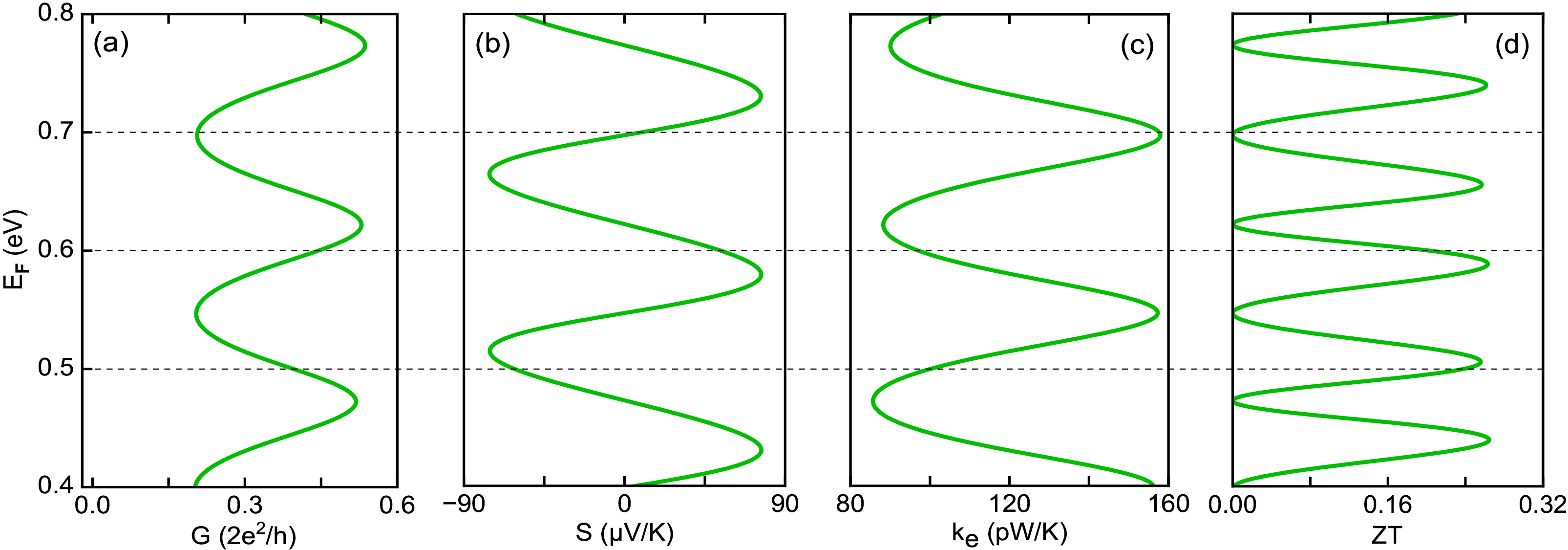}
    \caption{\small(Color online). (a) Electrical conductance ($G$), (b) Seebeck coefficient ($S$), (c) thermal conductance ($k_e$), and (d) figure of merit ($ZT$) as a function of Fermi energy ($E_F$) for $\nu=0\,$ with $N=40\,$, $T=200\,$K, and $W=1\,$eV.}
    \label{fig.4}
\end{figure*}
\begin{figure*}[htbp]
\includegraphics[width=16cm]{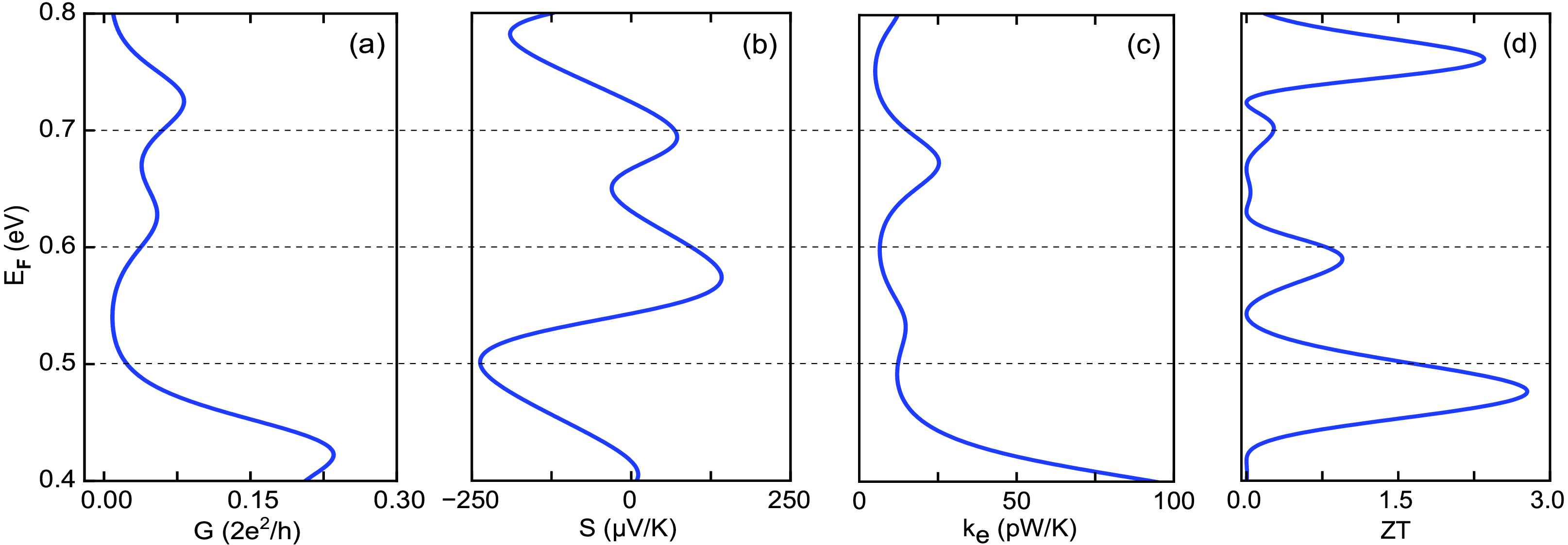}
\caption{\small(Color online). (a) Electrical conductance ($G$), (b) Seebeck coefficient ($S$), (c) thermal conductance ($k_e$), and (d) figure of merit ($ZT$) as a function of Fermi energy ($E_F$) for $\nu=0.55$ with $N=40$, $T=200\,$K, and $W=1\,$eV.}
\label{fig.5}
\end{figure*}
When the chain is disordered ($\nu=0.55\,$), the variation of the on-site potentials causes electron scattering, which induces localization
(Fig.~\ref{fig.3}(b)). So, transmission peaks are non-uniform for the entire energy window. Some large peaks are obtained along with some 
small ones. Moreover, this transmission spectrum looks asymmetric across any Fermi energy placed within the energy window of non-zero transmission. This irregular and asymmetric nature of the transmission profile is the primary requirement for getting high values of 
$ZT$. Placing the Fermi energy $E_F$, suitably, in this energy window, one can obtain a significantly large $ZT$.

Let us now discuss the thermoelectric quantities $G$, $S$, and $k_e$, and the energy conversion efficiency factor $ZT$ for both perfect and disordered cases (Figs.~\ref{fig.4} and~\ref{fig.5}). For $\nu=0\,$, i.e., when the chain is perfect, electronic wavefunctions are described by extended Bloch states. Due to the absence of scattering, electrons propagate freely throughout the medium, hence, transmission probability of electrons is high, which causes high values of electrical conductance $G$ (Fig.~\ref{fig.4}(a)).

In an ordered system, nearly uniform electronic transmission probability across the Fermi energy $E_F$ replicates equal contributions of electrons and holes, but in opposite sense to the thermopower $S$, which leads to mutual cancellation. This gives low values of $S$ as shown in Fig.~\ref{fig.4}(b). As the figure of merit ($ZT$) is proportional to $S^2$, even a small reduction in $S$ produces a moderate suppression in $ZT$ irrespective of how large $G$ is.

In a perfect lattice, electrons travel without experiencing resistive collisions, which indicates that heat transport is efficient over a wide energy spectrum. So, electronic thermal conductance $k_e$ is also high (Fig.~\ref{fig.4}(c)) in the absence of sharp variations in electronic transmission probability. Looking closely at the profiles of $G$ and $k_e$, one can identify that the values of $G$ and $k_e$ are no longer proportional to each other. For example, at $E_F=0.5\,$eV, $G$ and $k_e$ show exactly opposite nature, where $G$ is increasing but $k_e$ is decreasing. This nature is observed for the whole Fermi energy window. This is a clear indication of violating the Wiedemann-Franz law at the nanoscale regime, even at a perfect lattice. 

Due to smooth and uniform electronic transmission profile and low Seebeck coefficient, $ZT$ remains low for the entire Fermi energy window, as $S$ acts as a limiting factor, which can be seen from Fig.~\ref{fig.4}(d).

Now, let us explain all the quantities for the disordered case i.e., for $\nu=0.55\,$. We will start the discussion with the electrical conductance $G$. Figure~\ref{fig.5}(a) displays how $G$ varies with Fermi energy $E_F$, where $G$ is much smaller (close to zero) at some Fermi energies and at other energies it also remains a little higher. When we impose correlated disorder into the system, electronic transmission function {\Large$\tau$}($E$) becomes irregular with some large peaks and some small peaks, as shown in Fig.~\ref{fig.3}(b). As $G$ is directly proportional to {\Large$\tau$}($E$), it becomes a little higher when the transmission peak is high, and it gets much smaller when the transmission peak is small. But, overall $G$ is much lower than the perfect case (Fig.~\ref{fig.4}(a). At some Fermi energies, we obtain a high $S$ even while $G$ gets small. (Fig.~\ref{fig.5}(b)).

Now, we arguably discuss the logic of getting a high $S$ value. In Fig.~\ref{fig.5}(b), variation of Seebeck coefficient $S$ is shown as a function of Fermi energy. For some Fermi energies, the value of $S$ is increased significantly compared to the perfect case (Fig.~\ref{fig.4}(b)), while for the other energies, though the increase is not so high, but still favorable for getting a large value of $ZT$. As $ZT$ is proportional to the square of $S$, a small increase in the value of $S$ gives a large $ZT$. As $S$ depends on the non-uniformity of {\Large$\tau$}($E$) across the Fermi energy, a more non-uniform {\Large$\tau$}($E$) gives larger $S$. Due to imposed correlated disorder, the electronic transmission probability {\Large$\tau$}($E$) becomes non-uniform (Fig.~\ref{fig.3}(b)) over the entire energy window. Placing the $E_F$ suitably in the energy window of non-zero transmission one can obtain a high value of $S$. As $S$ appears as a square term in the expression of $ZT$, the sign of $S$ does not matter. 
\begin{figure}[htbp]
	\centering
	\includegraphics[width=0.65\linewidth]{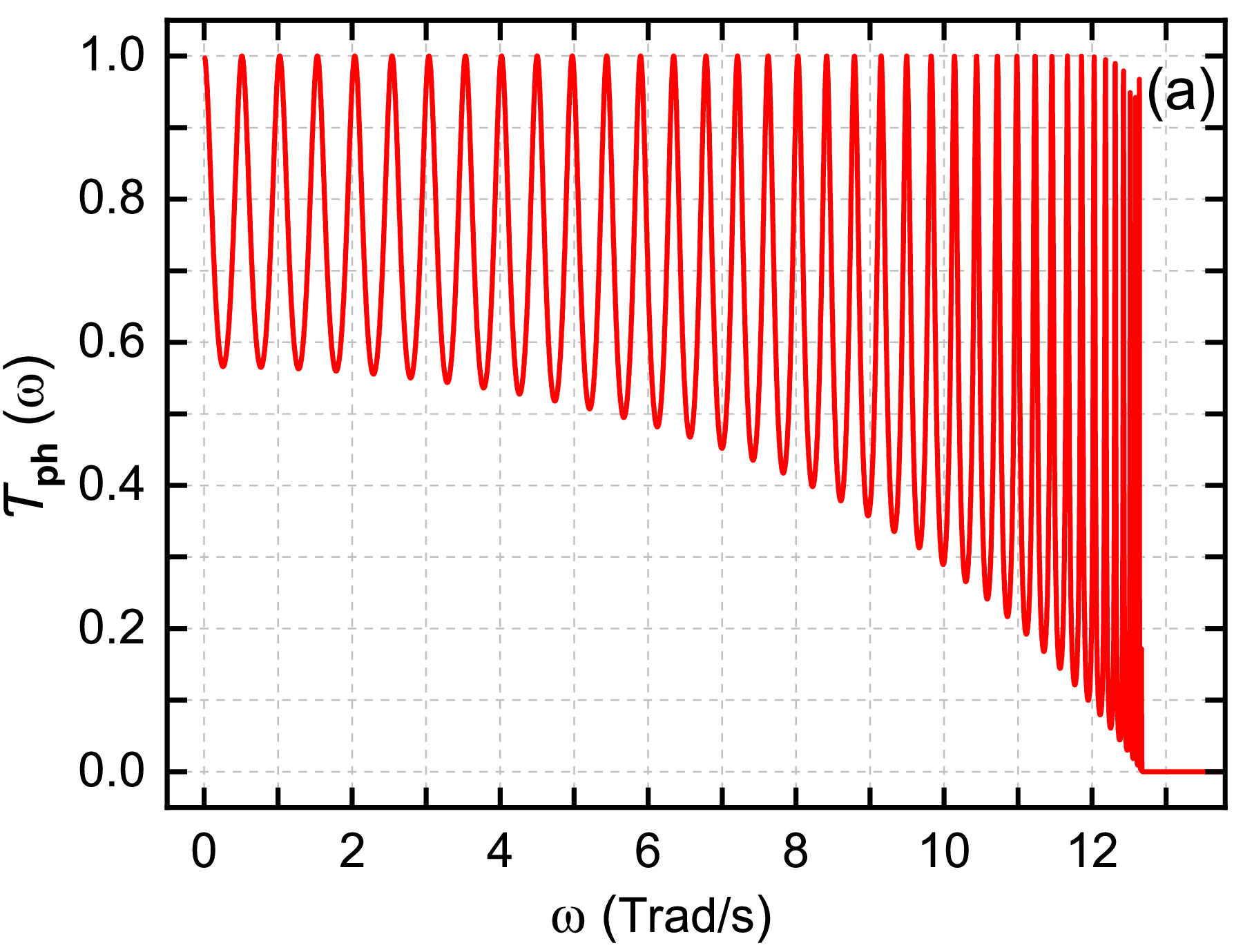}\vskip\baselineskip
	\includegraphics[width=0.65\linewidth]{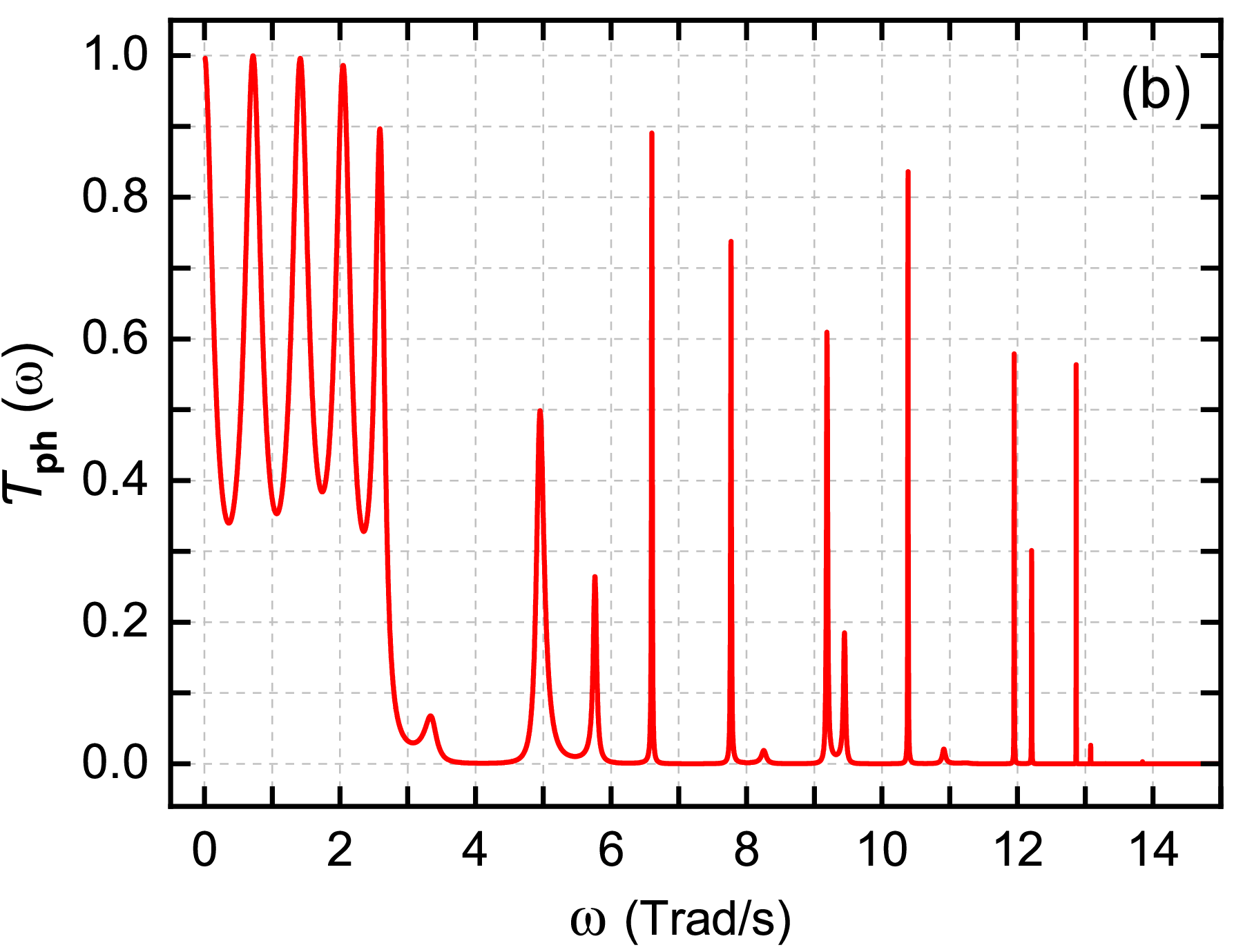}\vskip\baselineskip
	\includegraphics[width=0.65\linewidth]{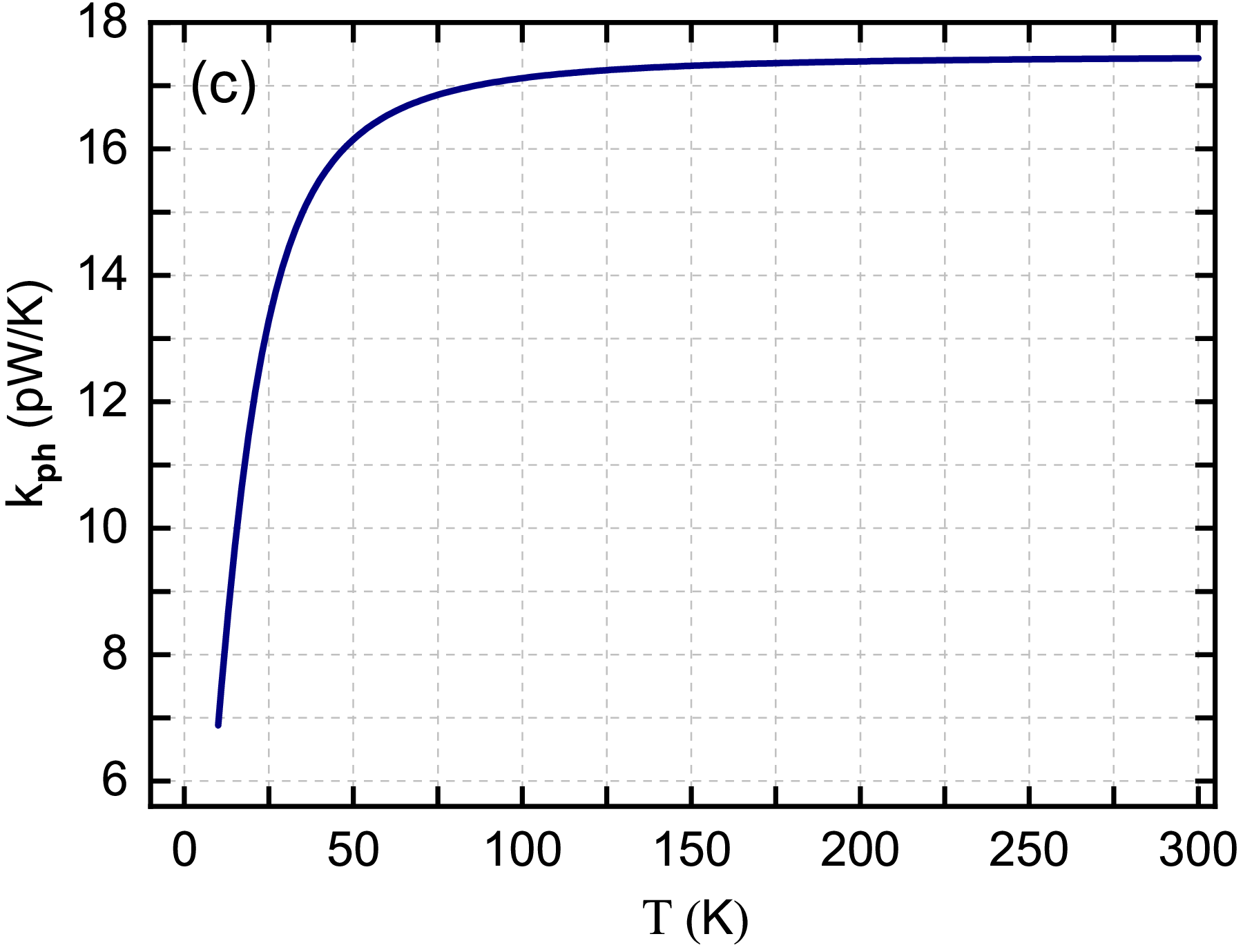}\vskip\baselineskip
	\includegraphics[width=0.65\linewidth]{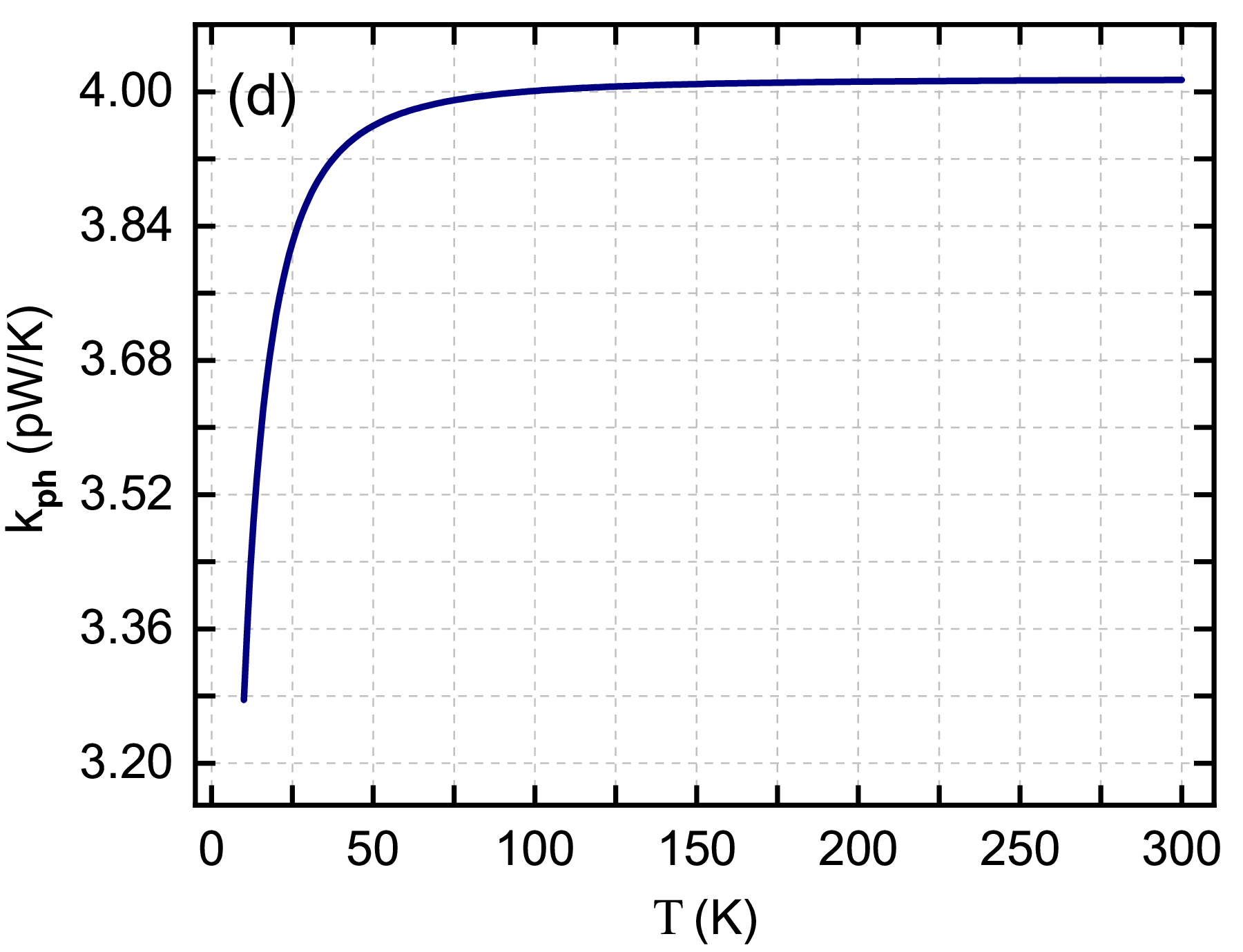}
	\caption{(Color online). Phonon transmission probability as a function of phonon frequency for (a) perfect case ($\nu=0\,$) and (b) disordered case ($\nu=0.55\,$). Phonon thermal conductance as a function of temperature for (c) perfect case ($\nu=0\,$) and (d) disordered case ($\nu=0.55\,$). All the figures are shown for set-1 with $N=40\,$ and $W=1\,$eV.}
	\label{fig.6}
\end{figure}
If {\Large$\tau$}($E$) is zero over any energy window around $E_F$, then $L_0$ becomes zero. If $L_0$ is zero at any $E_F$, $S$ diverges at that Fermi energy as $L_0$ appears at the denominator in the expression of $S$, which is nonphysical. So, one must be cautious in the calculation of the Seebeck coefficient. 

Another important factor for determining $ZT$ is electronic thermal conductance $k_e$. 
\begin{figure}[htbp]
	\centering
	\includegraphics[width=0.65\linewidth]{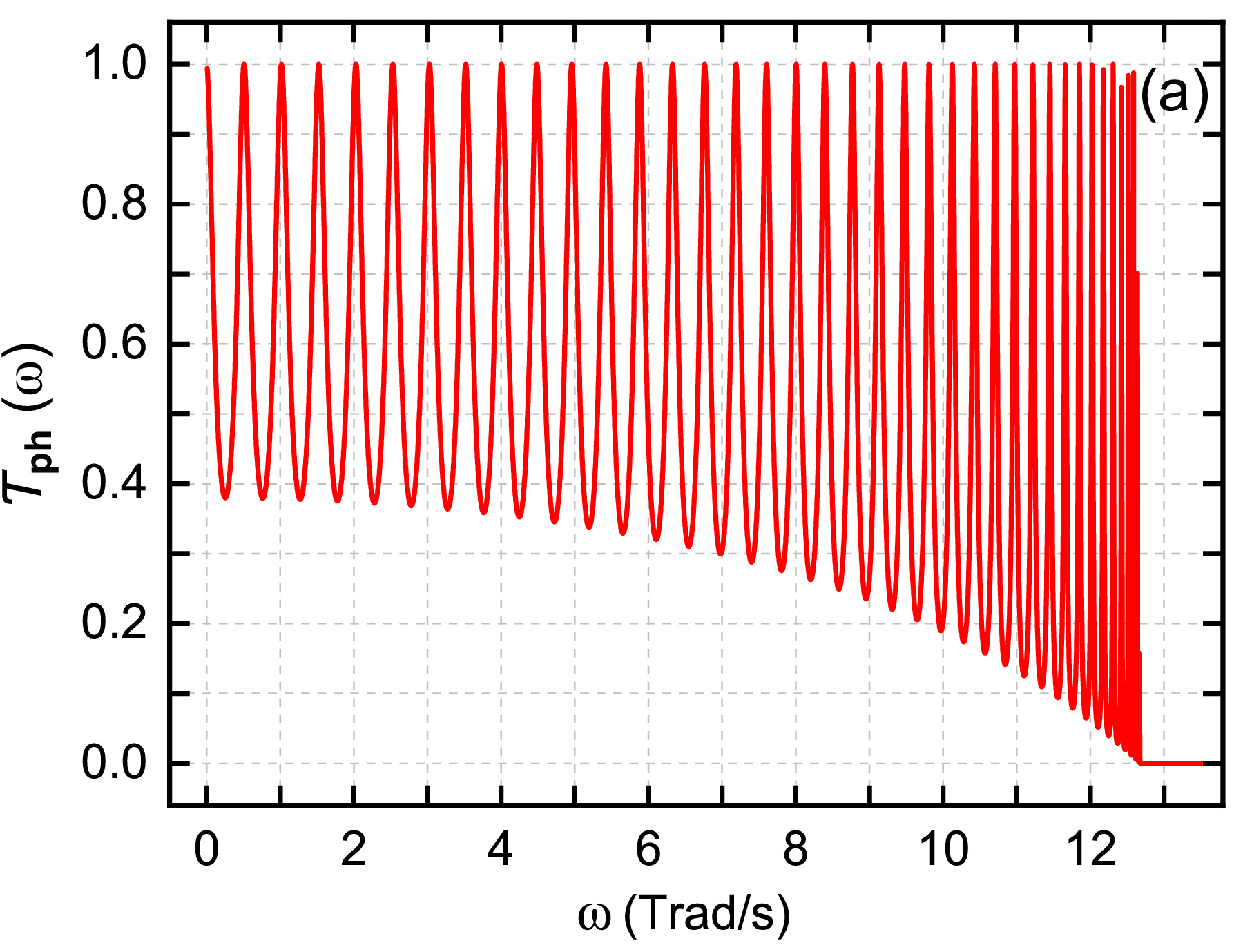}\vskip\baselineskip
	\includegraphics[width=0.65\linewidth]{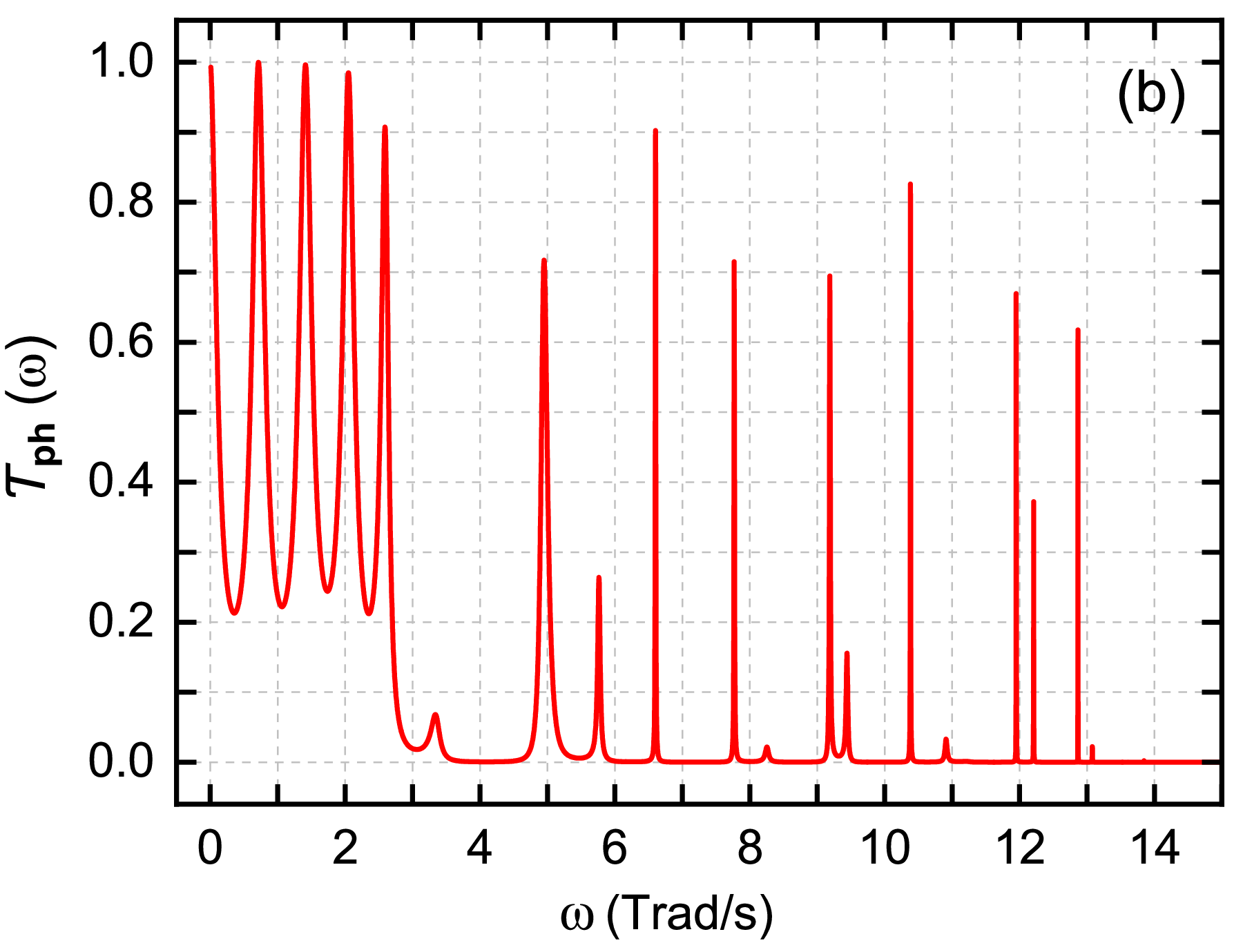}
	\vskip\baselineskip
	\includegraphics[width=0.65\linewidth]{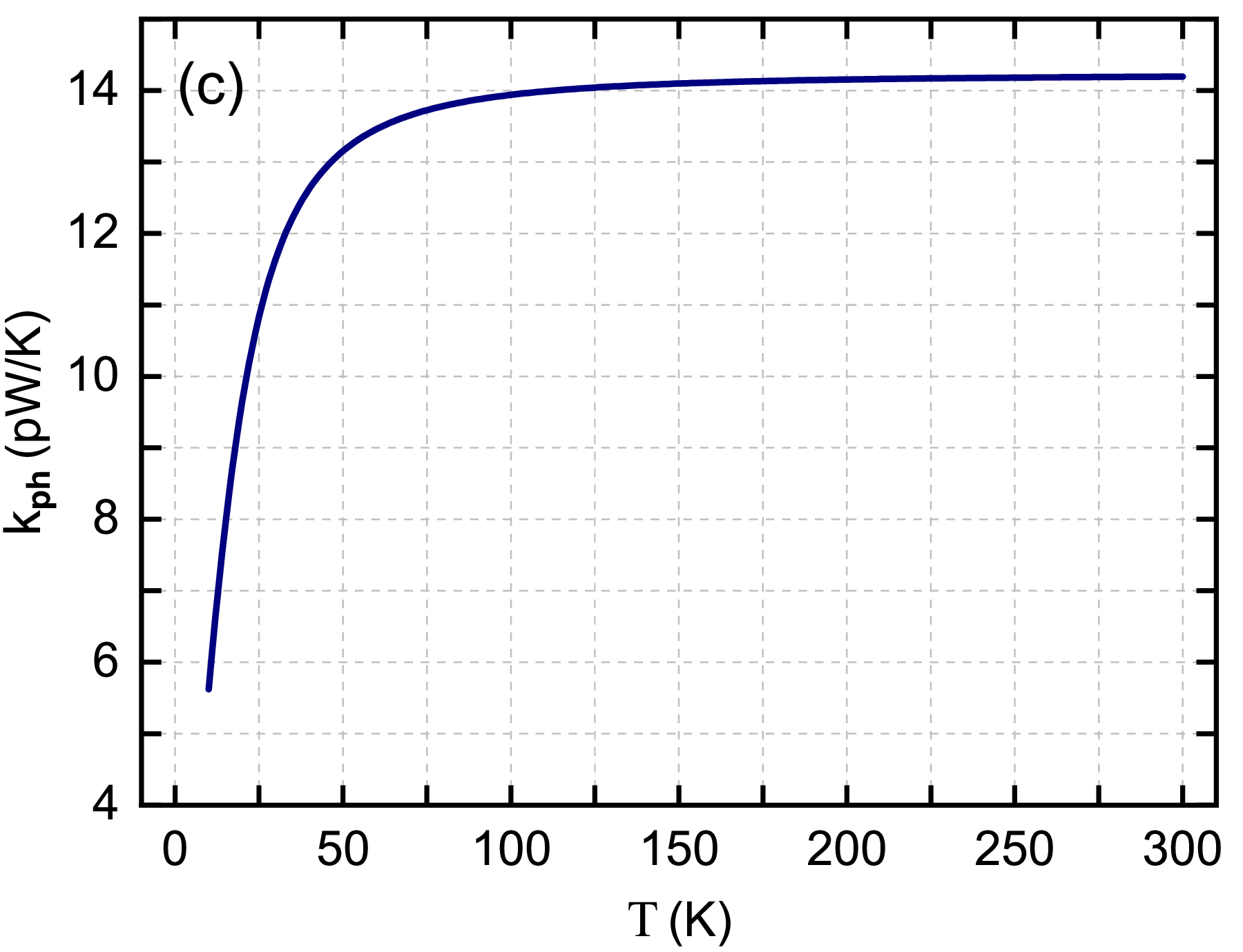}\vskip\baselineskip
	\includegraphics[width=0.65\linewidth]{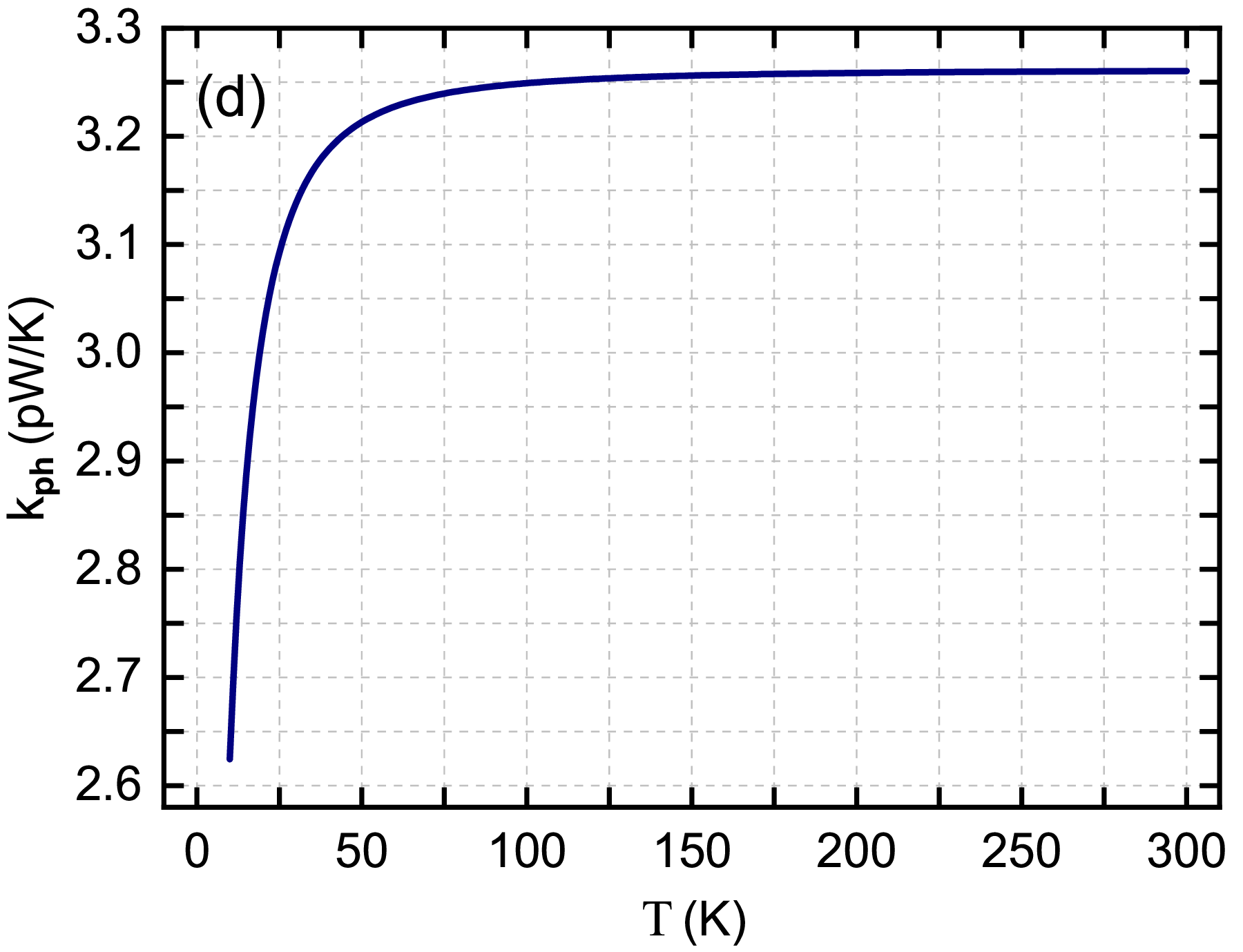}
	\caption{\small(Color online). Phonon transmission probability as a function of phonon frequency for (a) perfect case ($\nu=0\,$) and (b) disordered case ($\nu=0.55\,$). Phonon thermal conductance as a function of temperature for (c) perfect case ($\nu=0\,$) and (d) disordered case ($\nu=0.55\,$). All the figures are shown for set-2 with $N=40\,$ and $W=1\,$eV.}
	\label{fig.7}
\end{figure}
Finally, we concentrate on the variation of $k_e$ with Fermi energy (Fig.~\ref{fig.5}(c)). As $k_e$ comes at the denominator in the formulation of $ZT$, reduction of $k_e$ is also necessary for increasing the value of $ZT$. So, a small value of $k_e$ is required for achieving a favorable $ZT$. Non-zero values of electronic thermal conductance $k_e$ are obtained at different Fermi energies similar to electrical conductance $G$. Although $G$ and $k_e$ are naturewise almost identical, like perfect case ($\nu=0\,$), they are no longer dependent on each other. 
So, the Wiedemann-Franz law remains invalid when we impose disorder. 

After discussing $G$, $S$, and $k_e$, now we will focus on the behavior of $ZT$ with the Fermi energy $E_F$. 
High values of $ZT$ ($>$ 1) are obtained at multiple Fermi energies, with a maximum value reaching almost 3 (Fig.~\ref{fig.5}(d)). Such values are consistent with large values of $S$ and low $k_e$, and it can be verified by looking at the spectra of $S$ (Fig.~\ref{fig.5}(b)) and $k_e$ (Fig.~\ref{fig.5}(c)). For example, from Fig.~\ref{fig.5} it can be viewed that at $E_F=0.4\,$eV, $k_e$ is large, but $S$ becomes nearly zero with finite $G$. So, $ZT$ becomes almost zero at this Fermi energy. Similarly, at $E_F=0.5\,$eV in spite of having a lower $G$, we get a good $ZT$ due to large $S$ and low $k_e$.

\vskip 0.1cm
\noindent
\textbullet\ \textbf{Contribution of phonon to thermoelectric efficiency}: In this part, we elaborately discuss about the contribution 
of phonon for carrying out a precise calculation of $ZT$. For this we choose two different sets of systems, where we describe each system 
as a spring-mass system. The sets differ in terms of the masses and spring constants of the system and the side-coupled contacts, and 
the reason for considering different sets is to check how sensitive the results are to these parameters.

(a) \underline{Set-1}: It is assumed that the source and drain are made of the same material. The atomic mass and spring constant for 
the source/drain are taken as $4.7\times10^{-26}\,$kg and $16.9\,$N/m, respectively. For the central chain, these parameters are 
$1.99\times10^{-26}\,$kg and $1.6\,$N/m, respectively. By averaging the atomic masses and spring constants of the contacts and the 
chain, the cutoff frequency is obtained as $\omega_c=33.2\,$Trad/s. The values of all these parameters are consistent with the values 
reported in the literature~\cite{56}.

(b) \underline{Set-2}: This is another setup in which we assume that the source-channel-drain junction is made of different materials 
than what are considered in set-1. The atomic mass and the spring constant for the source/drain are taken to be $1.2\times10^{-25}\,$kg 
and $13.7\,$N/m, respectively. For the chain, these values are $1.99\times10^{-26}\,$kg and $1.6\,$N/m, respectively. By averaging the 
atomic masses and spring constants of the source/drain and chain, the cutoff frequency is obtained as $\omega_c=20.9\,$Trad/s.

Similar to the choice of site energies in the weakly varying disordered case, for the spring-mass system we introduce disorder in the 
masses~\cite{massmod} through a cosine modulation of the form $M_n = M_0\left[1 + W\cos(2\pi n^\nu)\right]$, where $W$ and $\nu$ denote 
the disorder 
strength and the incommensurate factor, respectively. Here, $M_0$ represents the uniform mass. For $\nu = 0$ and $1$, the system becomes
perfectly ordered, whereas for all other cases ($0 < \nu < 1$) it remains disordered, similar to the situation considered in the 
electronic case.

In Fig.~\ref{fig.6}(a) and Fig.~\ref{fig.7}(a), the variation of phonon transmission probability {\Large{$\tau_{\scriptscriptstyle ph}$}} is shown as a function of phonon frequency $\omega$ for set-1 and set-2, respectively. For perfect case, several Fabry-Pérot-like peaks\cite{56} are seen in both figures. At low frequencies, the phonon wavelength is much higher than the interatomic spacing. So, phonons propagate coherently like continuum elastic waves, suffering negligible reflection at the atomic sites. So, the phonon transmission probability is very high and close to unity. For higher frequencies, the phonon wavelength becomes comparable to the lattice spacing, resulting in enhanced scattering, which reduces the transmission probability.

In a disordered lattice (Fig.~\ref{fig.6}(b) and Fig.~\ref{fig.7}(b)), phonon transmission profiles become 
\begin{figure}[htbp]
	\includegraphics[width=6cm]{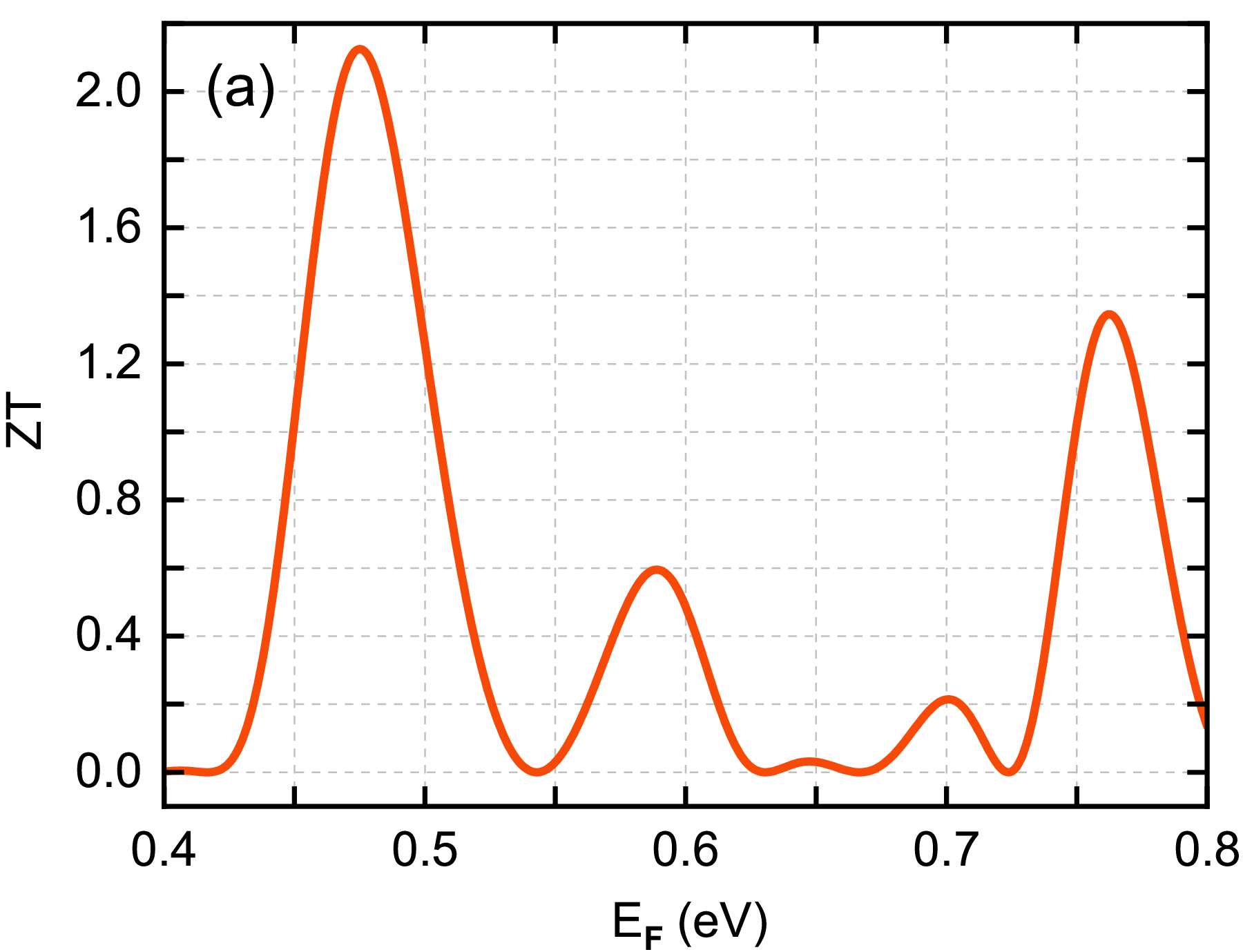} 
	\vskip\baselineskip
	\includegraphics[width=6cm]{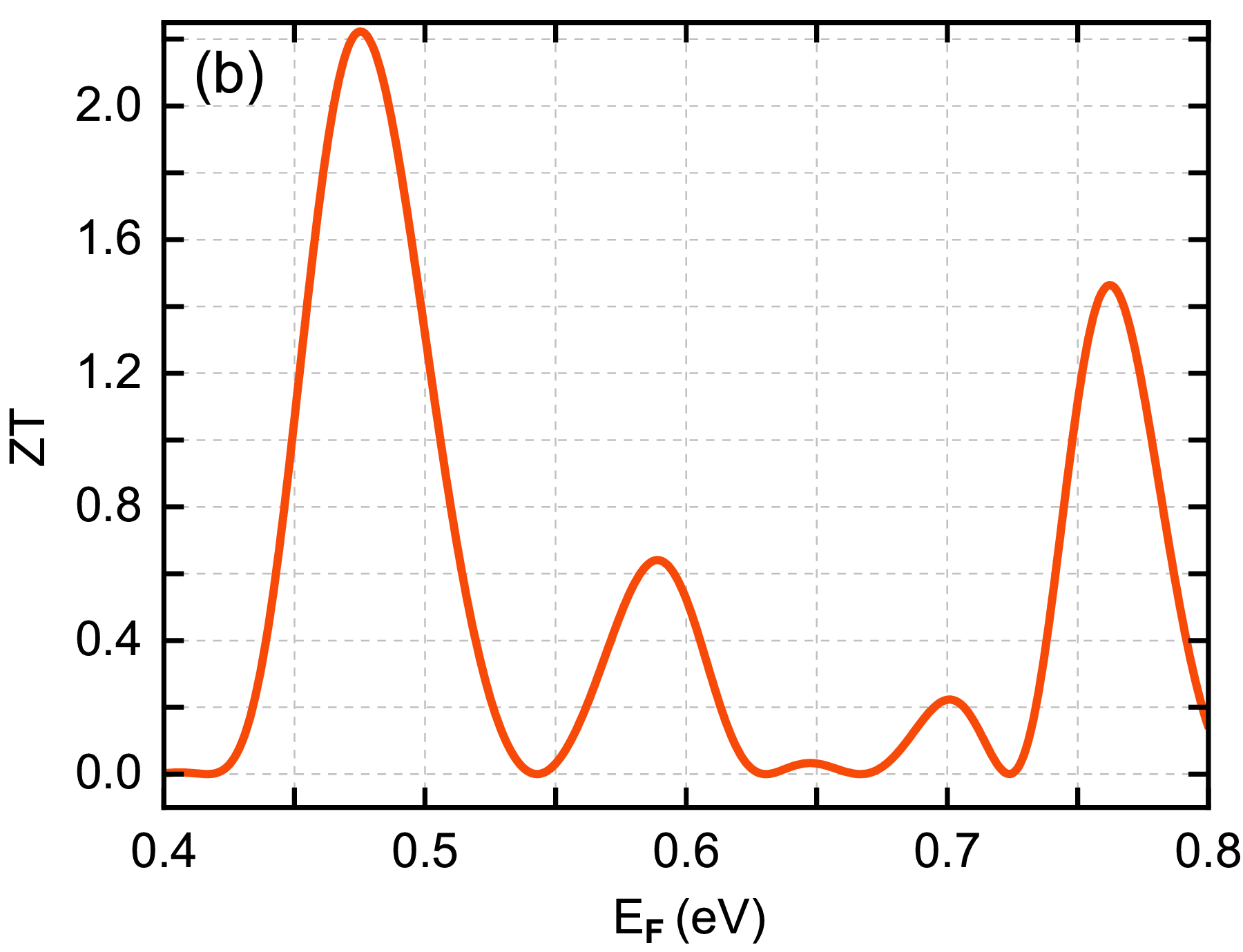} 
	\caption{\small(Color online). $ZT$ as a function of Fermi energy ($E_F$) for (a) set-1 and (b) set-2 with $\nu=0.55\,$, $N=40\,$, $W=1\,$eV, and $T=200\,$K, taking the effect of both electronic and phonon thermal conductances.} 
	\label{fig.8}
\end{figure}
irregular in nature with 
some discrete peaks within the entire frequency window for both systems because disorder destroys translational symmetry and induces 
backscattering. 

\begin{figure*}[ht]
	\centering
	\includegraphics[width=0.4\linewidth]{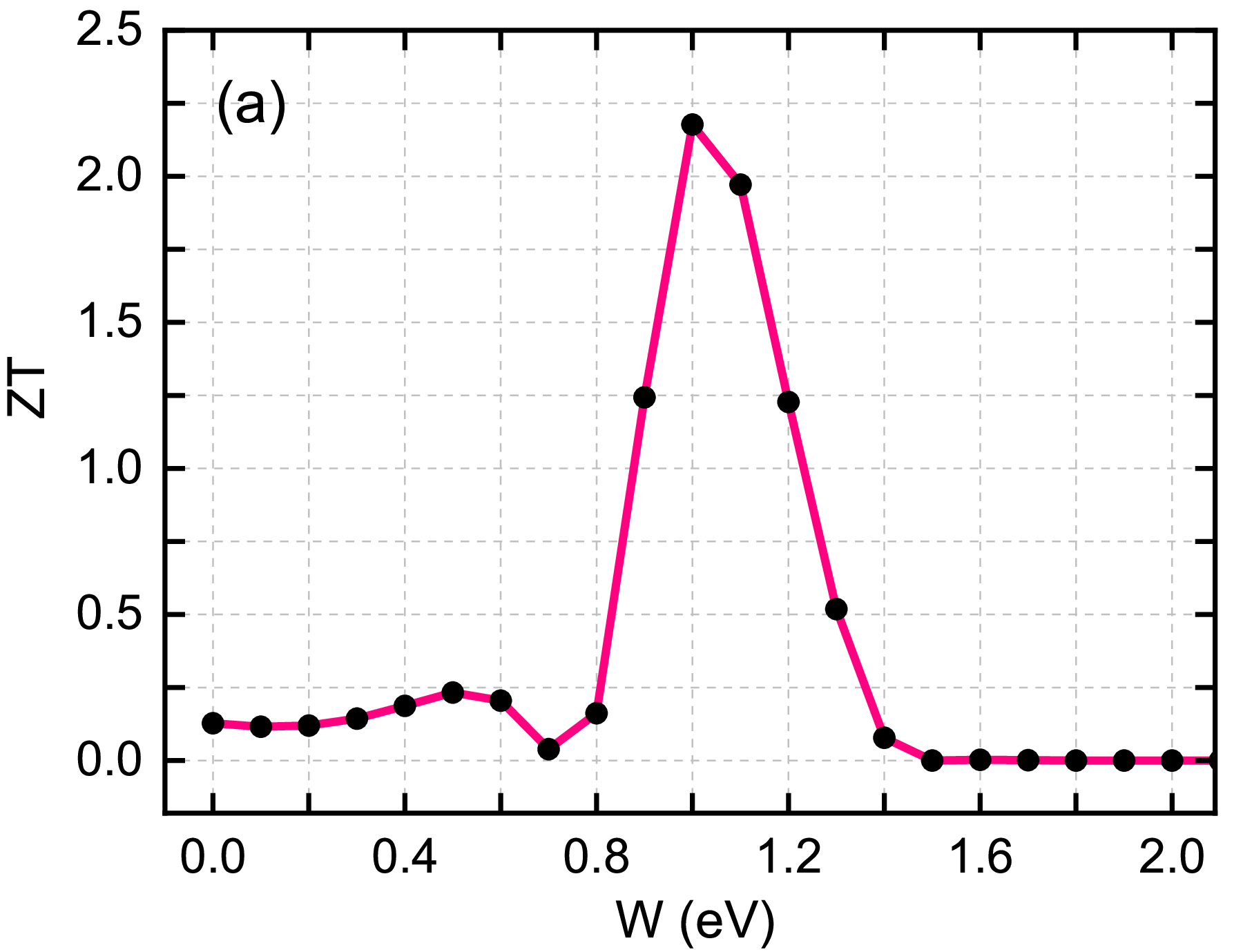}\hspace{10pt}
	\includegraphics[width=0.4\linewidth]{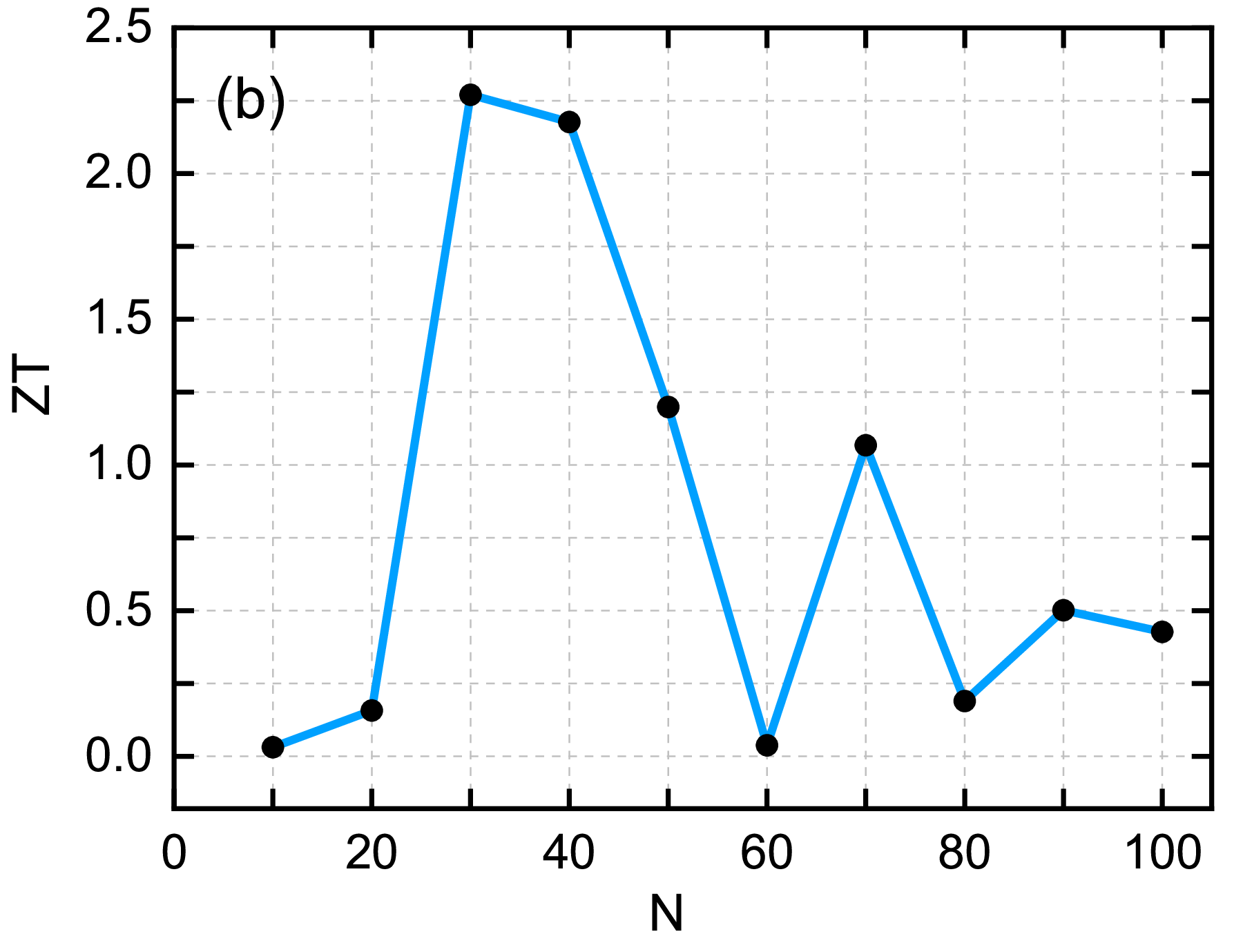}
	\vskip\baselineskip
	\includegraphics[width=0.4\linewidth]{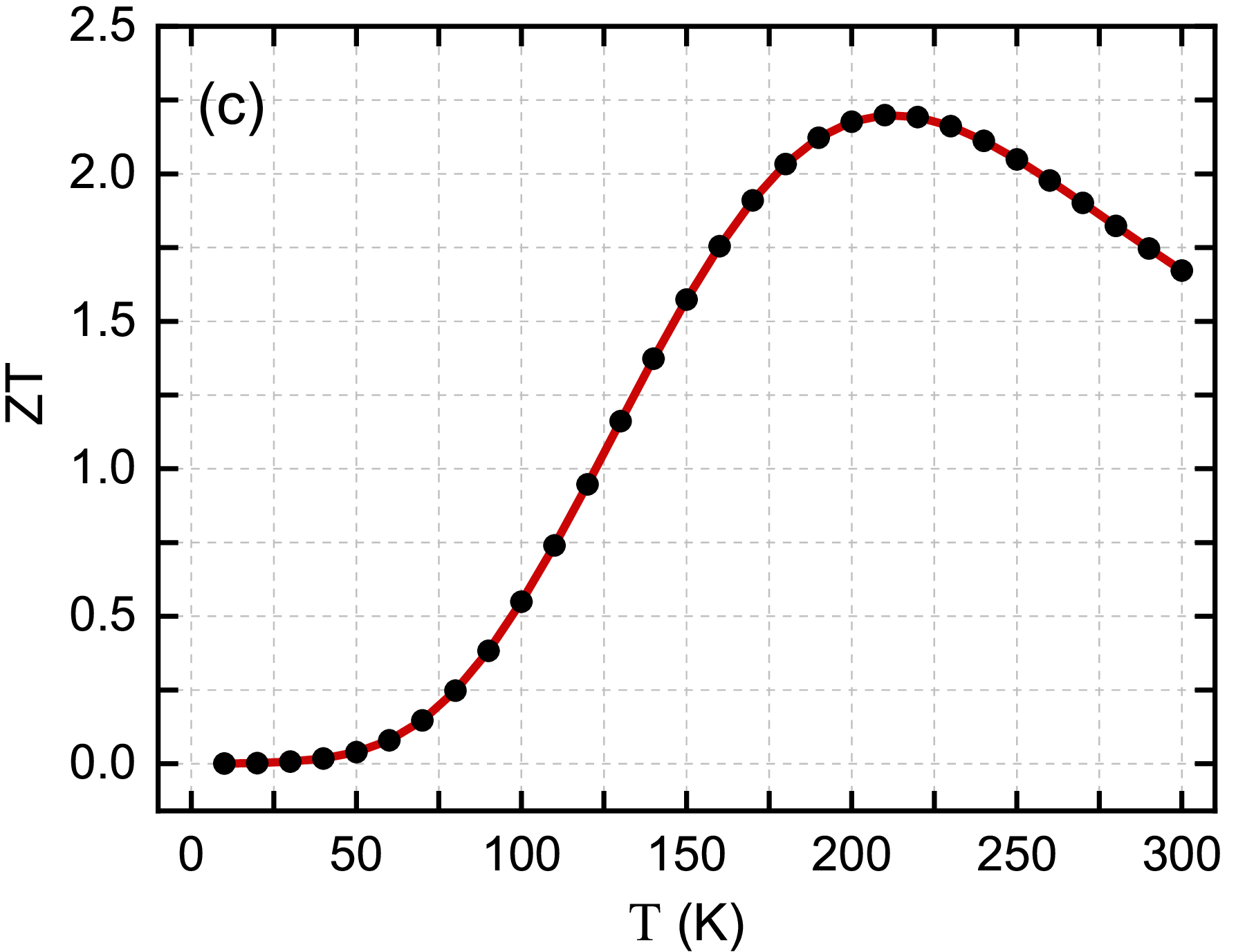}\hspace{10pt}
	\includegraphics[width=0.4\linewidth]{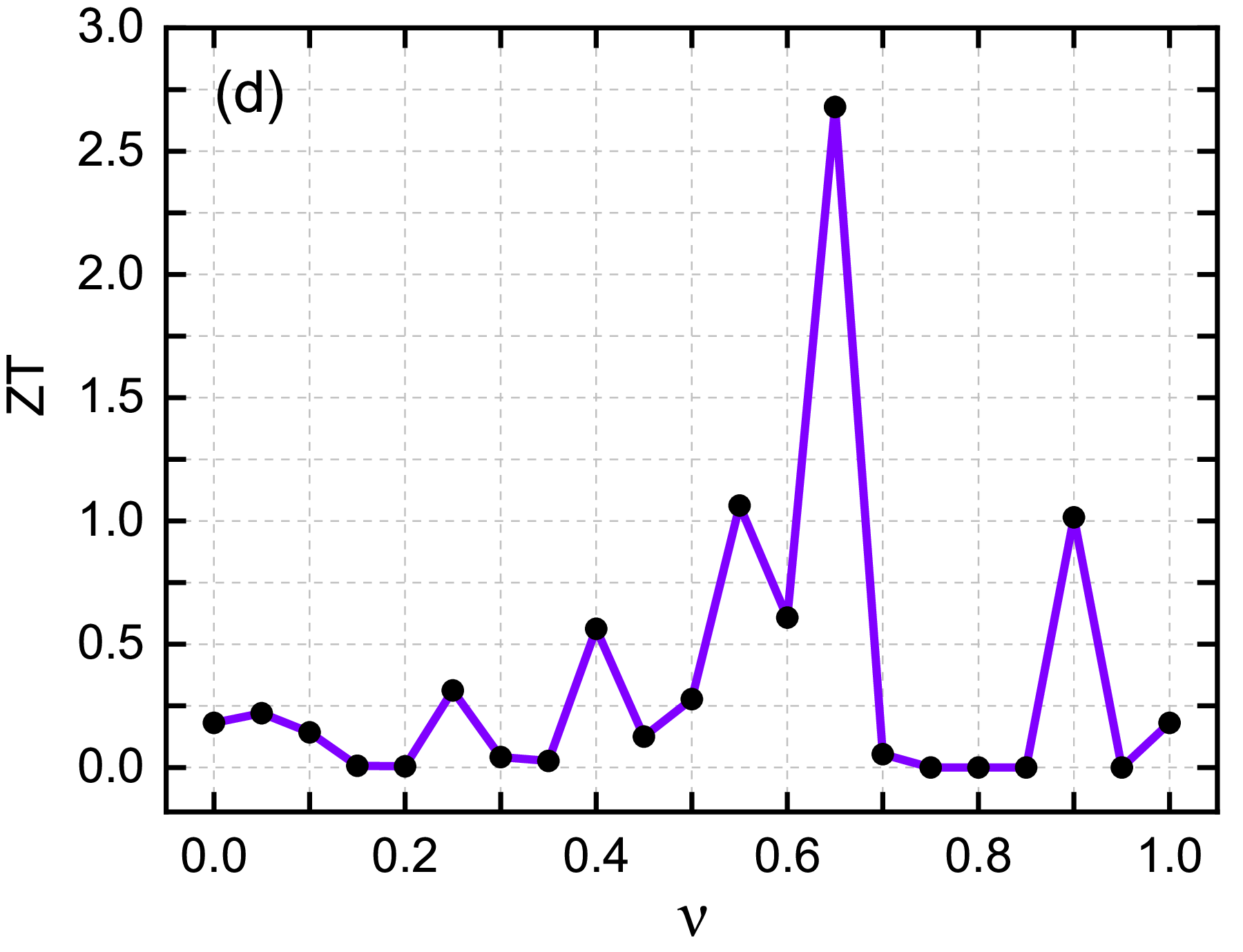}
	\caption{\small (Color online). Variation of $ZT$ with (a) disorder strength ($W$) for $E_F=0.48\,$eV, $\nu=0.55\,$, $T=200\,$K, and 
		$N = 40\,$; (b) chain size ($N$) for $E_F=0.48\,$eV, $\nu=0.55\,$, $T=200\,$K, and $W=1\,$eV; (c) temperature ($T$) for $E_F=0.48\,$eV,
		$\nu=0.55\,$, $W=1\,$eV, and $N=40\,$; and (d) incommensurate factor ($\nu$) for $E_F = 0.45\,$eV, $W=1\,$eV, $N=40\,$, and $T=200\,$K.
		All the results are worked out for set-2.}
	\label{fig.9}
\end{figure*}

In Fig.~\ref{fig.6}(c) and Fig.~\ref{fig.7}(c), variation of phonon thermal conductance $k_{ph}$ as a function of temperature is shown 
for set-1 and set-2, respectively, for the perfect case. At low temperature, phonon thermal conductance rises rapidly within the 
temperature range $10-100\,$K, reaching saturation values $17.4\,$pW/K (Fig.~\ref{fig.6}(c)) and $14.1\,$pW/K (Fig.~\ref{fig.7}(c))
at high temperatures. These high saturation values are consistent with regular phonon transmission profiles obtained for the perfect cases.

For the disordered cases (Fig.~\ref{fig.6}(d) and Fig.~\ref{fig.7}(d)), same nature as perfect cases are observed with saturation values
reaching $4\,$pW/K and almost $3.3\,$pW/K for set-1 and set-2, respectively. These much lower saturation values than perfect cases are
consistent with irregular and non-uniform phonon transmission profiles. Increased back-scattering due to imposed correlated disorder 
reduces transmission; hence, phonon thermal conductance $k_{ph}$ decreases, as $k_{ph}$ is directly proportional to transmission 
probability {\Large{$\tau_{\scriptscriptstyle ph}$}}.

The results of $ZT$ discussed so far are worked out considering only the electronic part of thermal conductance. Now, we include both 
the electronic and phononic parts for a more accurate estimation of $ZT$. The results are given in Figs.~\ref{fig.8}(a) and (b) for 
the set-1 and set-2, respectively. The maximum value of $ZT$ reaches well above unity ($>2$) in both cases, which clearly emphasizes 
that our proposed system can be utilized for efficient energy conversion. The key message is that the inclusion of phonon thermal 
conductance does not reduce the efficiency factor appreciably. 

After discussing the contribution of phonon to $ZT$, here we focus on the behavior of $ZT$ with the other parameters, i.e., disorder 
strength $W$, chain size $N$, and temperature $T$, and we also study the effect of different configurations of this potential on 
$ZT$ by changing the incommensurate factor $\nu$. All the results are shown in Fig.~\ref{fig.9} for the set-2, as we get a more 
favorable thermoelectric response for this set compared to the other.

The variation of $ZT$ with disorder strength $W$ is critically investigated for $\nu=0.55\,$, in Fig.~\ref{fig.9}(a). Although $ZT$ is 
small at low values of $W$, it becomes high at some intermediate disorder strengths, with the maximum value reaching above $2$ for 
$W=1\,$eV. For high values of $W$, localization starts to dominate, hence both electrical and thermal conductances decrease sharply, 
resulting in vanishingly small values of $ZT$ at high disorder strengths. 

In Fig.~\ref{fig.9}(b), $ZT$ is plotted as a function of chain size $N$ for $\nu=0.55\,$ and $W=1\,$eV to examine the effect of system 
size on thermoelectric efficiency. High value of $ZT$ ($>1$) is observed for multiple values of $N$, with the maximum value reaching 
well above $2$ for $N=30\,$, which means our system shows efficient thermoelectric behavior over a wide range of system sizes.

The variation of $ZT$ with temperature is shown in Fig.~\ref{fig.9}(c), for $\nu=0.55\,$and $W=1\,$eV. As $ZT$ is directly proportional 
to the equilibrium temperature of the system $T$, initially it increases with temperature, reaches a maximum value exceeding $2$, 
and then it starts to decrease with the further increase of $T$. At higher temperatures, the sharpness of electronic transmission 
peaks diminishes due to thermal broadening, leading to a reduction in $ZT$. Although $ZT$ decreases at higher temperatures but 
still favorable from an energy conversion perspective. 

In Fig.~\ref{fig.9}(d), variation of the figure of merit $ZT$ as a function of the incommensurate factor ($\nu$) is shown for 
$E_F=0.45\,$eV. A zigzag nature is found, accompanied by reasonably large values of $ZT$ at multiple values of incommensurate factor 
$\nu$. Significantly high values of $ZT$ are found for some particular values of $\nu$ with a maximum magnitude reaching above $2.5$, 
which indicates that if we place the Fermi energy $E_F$ at $0.45\,$eV for these incommensurate values, highly asymmetric electronic 
\begin{figure}[htbp]
\includegraphics[width=6.5cm]{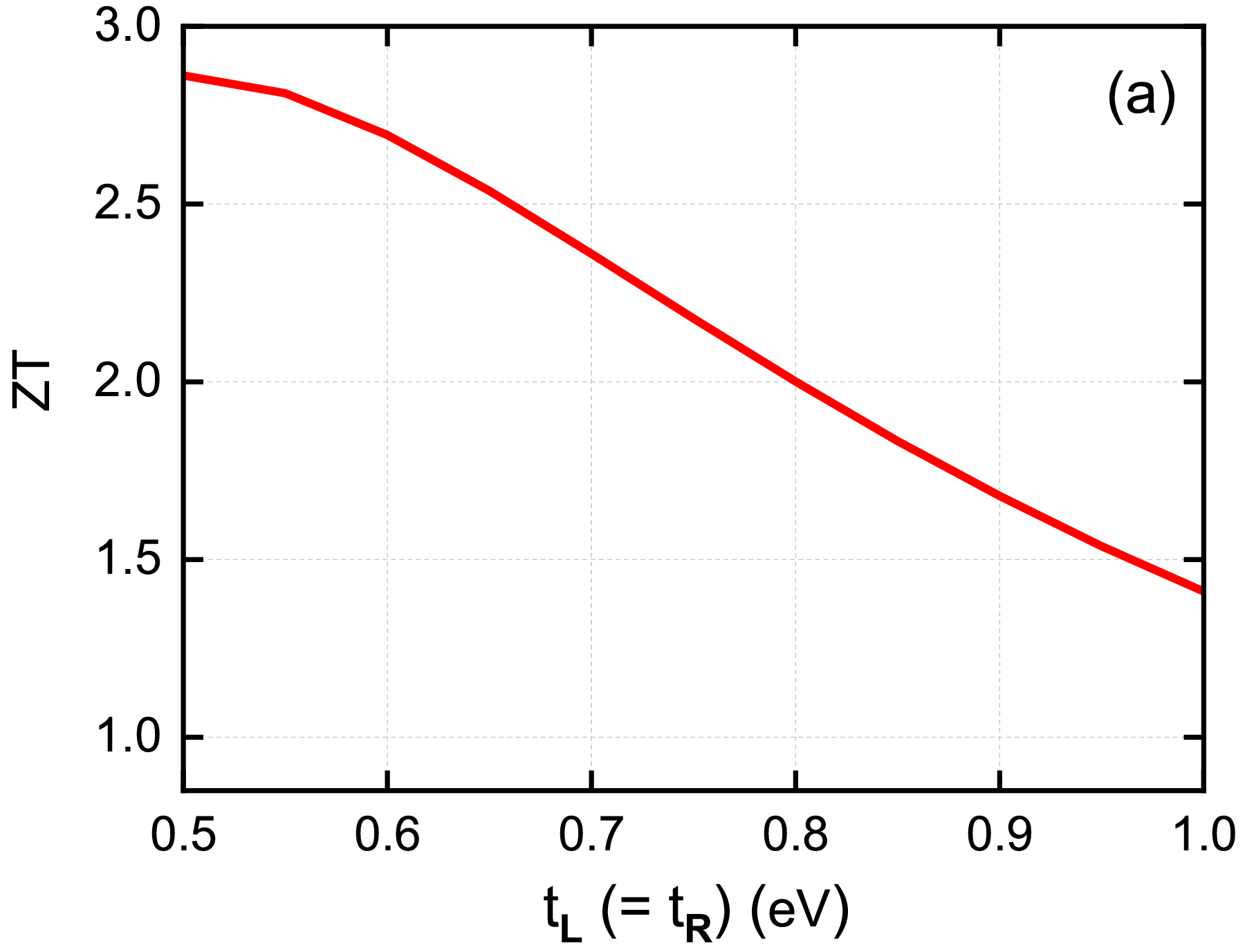} 
\vskip\baselineskip
\includegraphics[width=6.5cm]{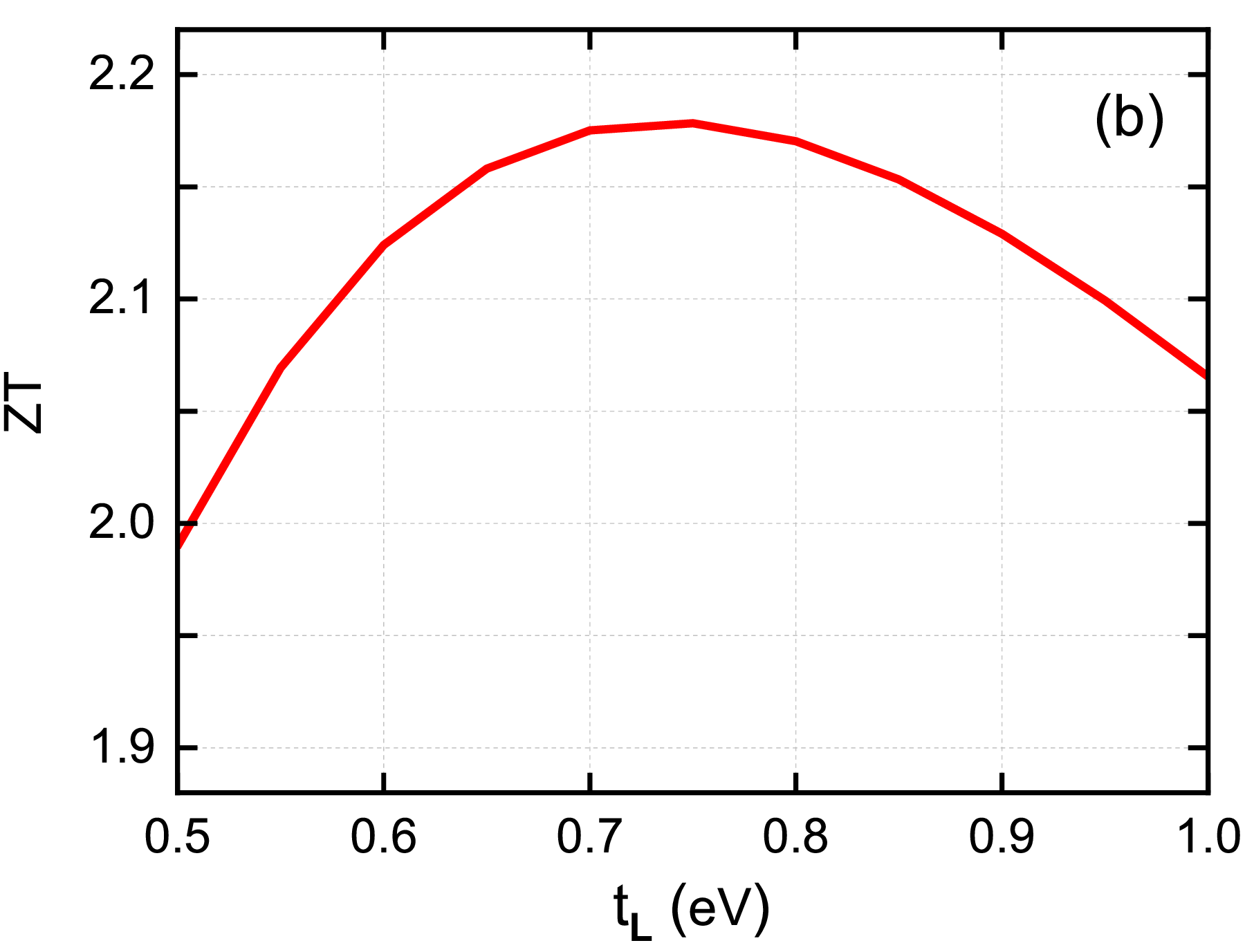} 
\caption{\small(Color online). $ZT$ as a function of (a) $t_L$, where both $t_L$ and $t_R$ vary; and (b) $t_L$, where 
$t_R$ is fixed at $0.75\,$eV. All other fixed parameters are: $E_F=0.48\,$eV, $\nu=0.55$, $N=40$, $W=1\,$eV, and $T=200\,$K. The results 
are worked out for set-2.}
\label{fig.10}
\end{figure}
transmission line-shapes are obtained, leading to high values of $ZT$. For $\nu=0\,$ and $\nu=1\,$, i.e., for perfect cases, $ZT$ values 
are identical, which is expected.

Now, we discuss the impact of coupling asymmetry on $ZT$, as chain to electrode coupling is important for describing thermoelectric performance. In Fig.~\ref{fig.10}(a), variation of $ZT$ is shown as a function of coupling parameter $t_L$, where we vary both the coupling strengths, and a monotonically decreasing nature of $ZT$ is obtained. As we increase the coupling strength $t_L$, $ZT$ decreases, and relatively smaller values are obtained at higher coupling strengths. Although a decreasing nature is obtained, we observe a favorable response of $ZT$ over the entire coupling window, and moreover we can also say that value of $ZT$ can be enhanced by decreasing the coupling strength. On the other hand, when we vary $t_L$, keeping the coupling parameter $t_R$ fixed at $0.75\,$eV, a different nature of $ZT$ is obtained as shown in Fig.~\ref{fig.10}(b). Initially, $ZT$ increases with coupling parameter $t_L$, reaches a maximum value ($\sim2.2$), and again decreases. The maximum value of $ZT$ is obtained when $t_L$ and $t_R$ become equal. Although the nature of $ZT$ is different in Figs.~\ref{fig.10}(a) and (b), we observe a favorable response of $ZT$ over the entire coupling window in both figures. The change of
the coupling strengths directly affect the transmission profile, which therefore influences the energy conversion efficiency. The coupling 
effect on electronic transmission spectrum has been widely studied earlier, and hence, we do not discuss it explicitly in the present work. 
The results given in Fig.~\ref{fig.10} suggest the usefulness of our chosen model quantum system as an efficient thermoelectric energy 
conversion.

\begin{figure}[htbp]
\includegraphics[width=6.5cm]{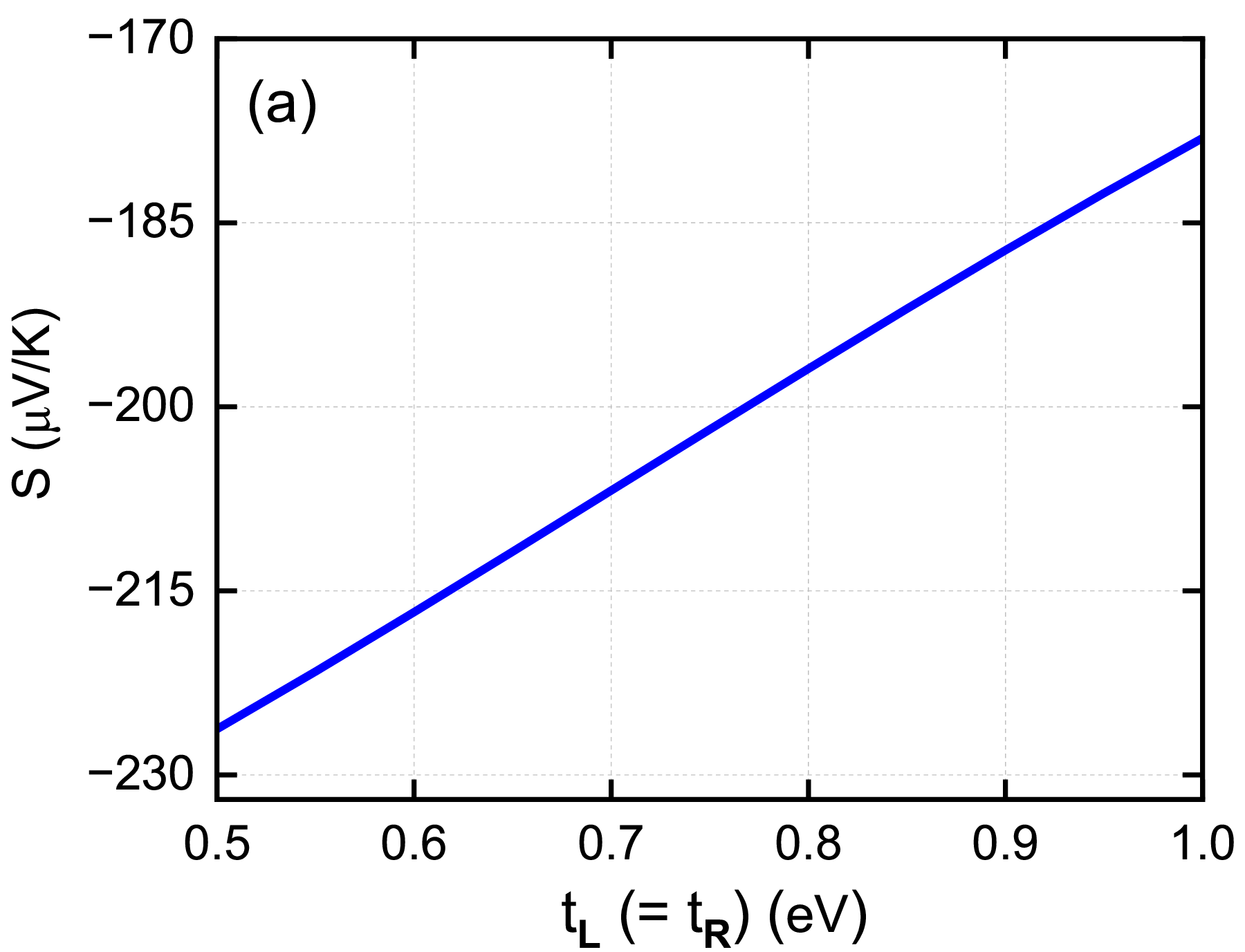} 
\vskip\baselineskip
\includegraphics[width=6.5cm]{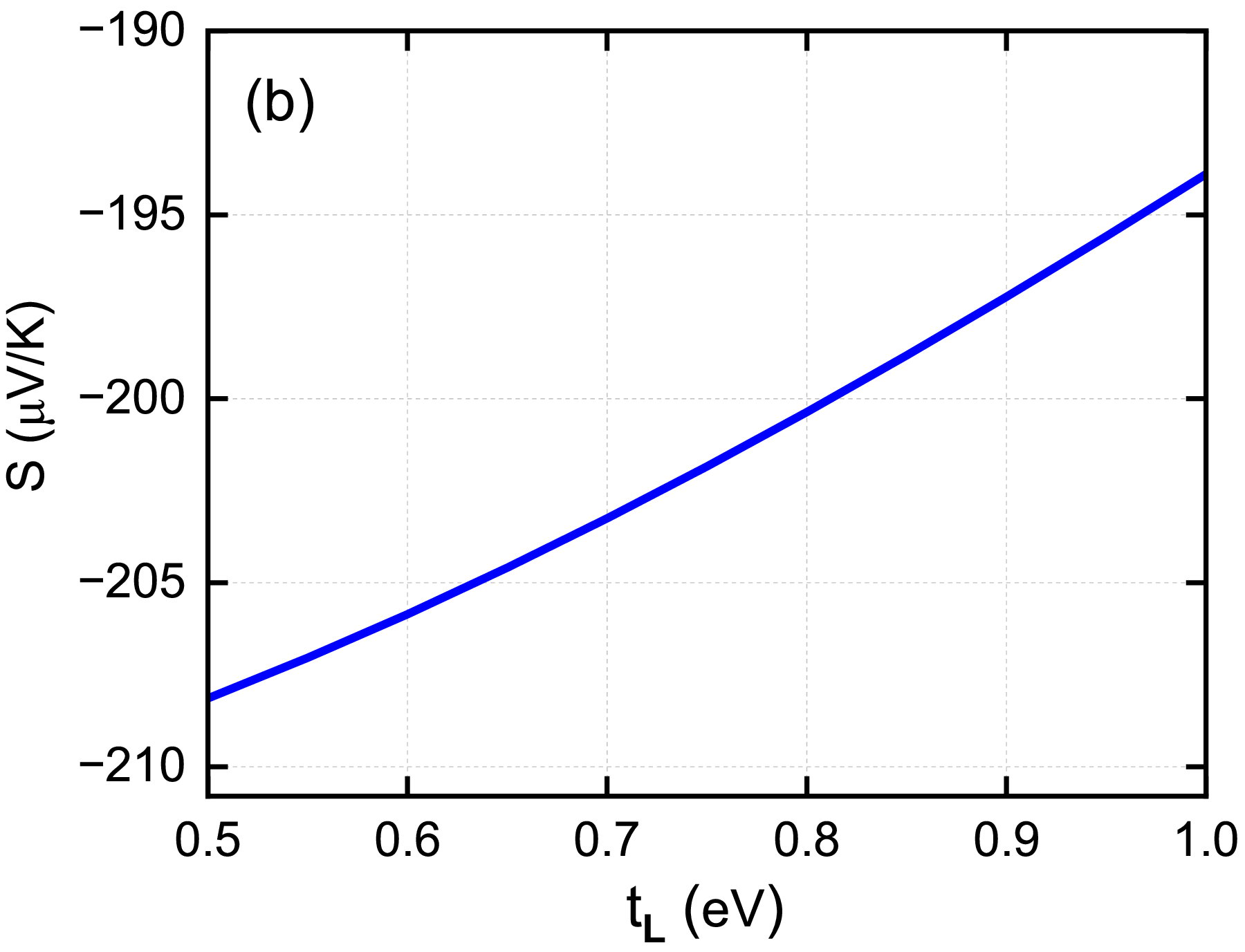} 
\caption{\small(Color online). Seebeck coefficient $S$ as a function of (a) $t_L$, where both $t_L$ and $t_R$ vary; 
and (b) $t_L$, where $t_R$ is fixed at $0.75\,$eV. All other fixed parameters are: $E_F=0.48\,$eV, $\nu=0.55\,$, $N=40\,$, $W=1\,$eV, 
and $T=200\,$K. The results are worked out for set-2.}
\label{fig.11}
\end{figure}

Seebeck coefficient $S$ is plotted as a function of the coupling parameters $t_L$ and $t_R$ in Fig.~\ref{fig.11},
to check the sensitivity of $S$ to the coupling strength. It is clearly seen that the magnitude of $S$ decreases as we 
increase the coupling strength $t_L$, shown in Fig.~\ref{fig.11}(a), which directly influences $ZT$ (Fig.~\ref{fig.10}(a)), where $ZT$ 
also decreases with enhancing the coupling strength. The magnitude of $S$ also gets reduced with the coupling strength $t_L$ when $t_R$ 
is kept fixed, as shown in Fig.~\ref{fig.11}(b), but the reduction is less compared to Fig.~\ref{fig.11}(a). The underlying mechanism 
relies on the modification of transmission lineshape with the coupling parameters.

\begin{figure}[htbp]
\includegraphics[width=7.5cm]{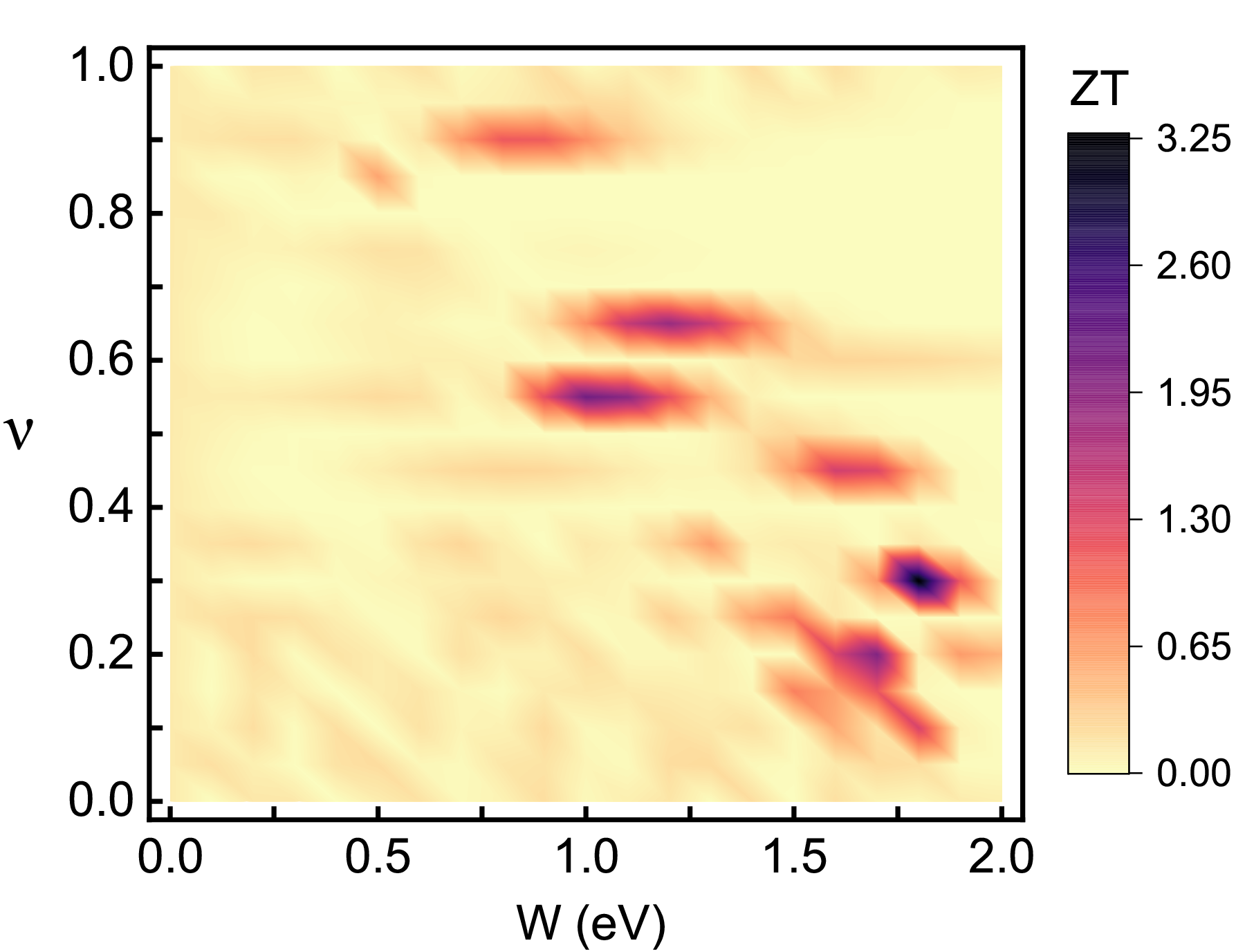}
\caption{\small(Color online). $ZT$ as a function of $W$ and $\nu$ with $E_F=0.48\,$eV, $N=40\,$, and $T=200\,$K for set-2.}
\label{fig.12}
\end{figure}

From our earlier discussion it is found that the thermoelectric efficiency $ZT$ is sensitive to the choices of the 
disorder strength $W$ and the incommensurate parameter $\nu$, as clearly reflected from the spectra given in Fig.~\ref{fig.9}(a) and
Fig.~\ref{fig.9}(d), respectively. To get a complete picture, we plot $ZT$ as a function of both $W$ and $\nu$ in Fig.~\ref{fig.12}, 
where the dark patches represent high values of $ZT$. Existence of many of these dark patches confirms that favorable values of $ZT$ 
are obtained for multiple values of disorder strengths ($W's$) and also for other configurations ($\nu's$) of the potential.
This density plot emphasizes that the favorable response is not specific to a particular set of parameter values, rather our results are 
valid for a broad range parameter values. Thus, we can claim the studied results can be checked in suitable laboratories.

From the above results it is found that for our chosen quantum system, the $ZT$ value reaches well-above the required 
limit ($ZT>1$). But, it is quite difficult to judge whether the high $ZT$ values correspond to practically useful thermoelectric 
performance or not. For that we need to calculate the power factor, output power, and the conductance at the optimal conditions.
The estimated values for the optimal parameter set ($E_F=0.48\,$eV, $W=1\,$eV, $T=200\,$K, and $\nu=0.55$) are as follows: 
Power factor = $GS^2=0.17\,pW/K^2$, Conductance $G$ (in units of $2e^2/h$) = $0.054$, and output power is $G(S\Delta T)^2=0.68\,$pW, 
with $\Delta T$ assumed to be $2\,$K. From these values, we claim that our proposed quantum system can safely be used for suitable 
energy conversion.

\vskip 0.25cm
\noindent 
$\blacksquare$ \textbf{Thermoelectric response of the conventional AAH system: Comparison with weakly varying AAH system}

\begin{figure}[htbp]
\includegraphics[width=7cm]{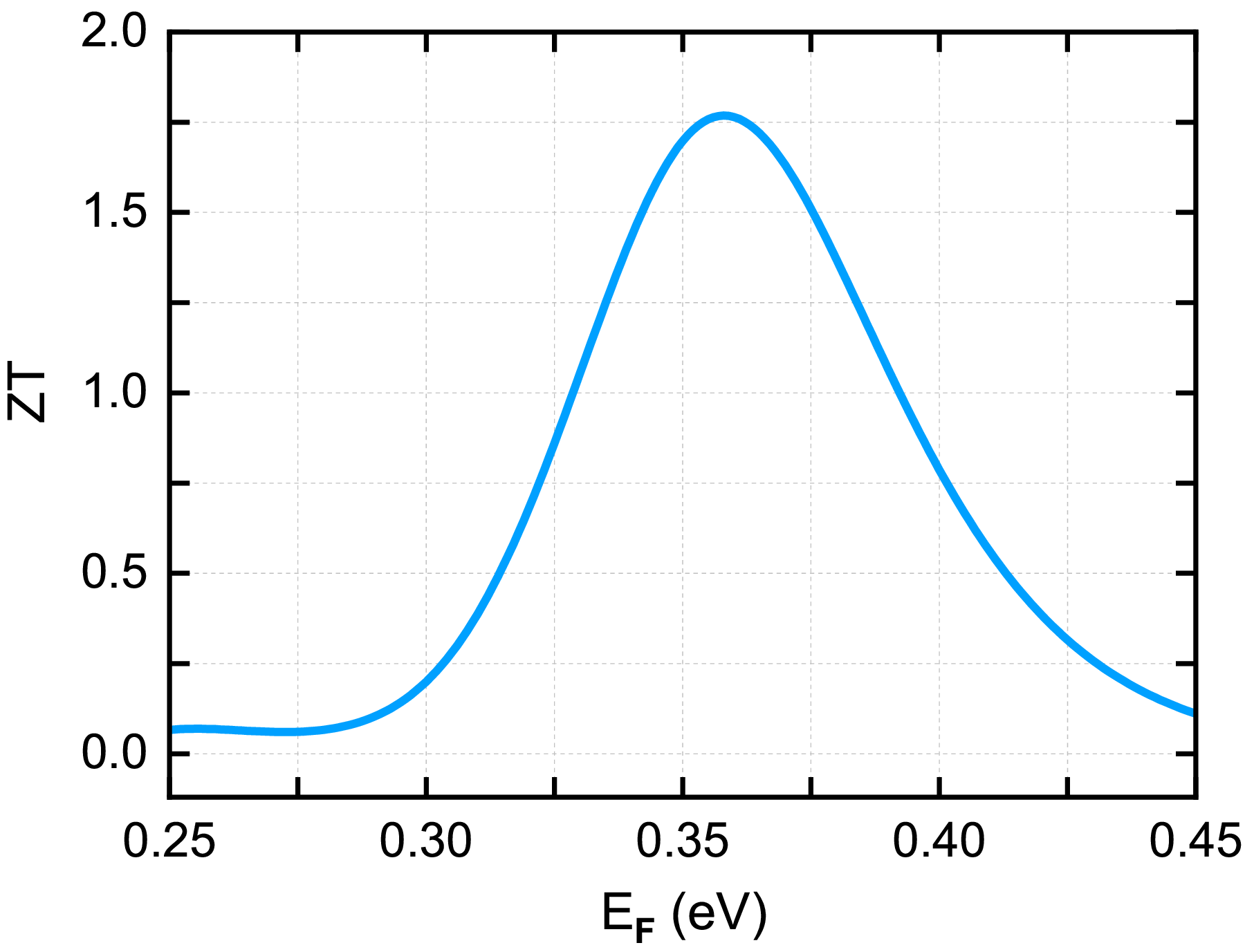}
\caption{\small(Color online). $ZT$ as a function of Fermi energy for $W=1\,$eV, $\phi=0$, $N=40\,$, and $T=200\,$K for the 
conventional AAH systems.}
\label{fig.13}
\end{figure}

\vskip 0.2cm 
So far, we have shown thermoelectric response for a weakly varying AAH system. However, it is important to compare 
the values of $ZT$ obtained for a weakly varying AAH system with those of any other correlated disordered system. So, here we consider a conventional AAH system, where site potentials are described as 
$\epsilon_n=W\cos(2\pi bn+\phi)$~\cite{34,45} with $W$ being the disorder strength. Here, $b$ is an irrational number and $\phi$ denotes 
the AAH phase. The key difference is that in weakly varying AAH systems, both localized and extended states exist simultaneously for 
different configurations of the potential, unlike the conventional AAH systems, where all states are either localized or extended 
depending on the disorder strength ($W$), as discussed in the introduction. Here, we present a result of $ZT$ for the conventional AAH 
system in Fig.~\ref{fig.13}, where we plot $ZT$ as a function of Fermi energy $E_F$ taking a $40$-site chain with $W=1\,$eV. Value of $b$ 
is set at $(1+\sqrt{5})/2$.
In Fig.~\ref{fig.13}, high values of $ZT$ are obtained at multiple Fermi energies. Although we obtain a 
favorable thermoelectric response in this case, the $ZT$ values found here are comparatively lower than those of the weakly varying 
AAH system. So, we can claim that the weakly varying AAH systems to be more thermoelectrically efficient than the conventional 
AAH systems.

Recent studies have demonstrated that strain engineering and site-specific atomic substitution can suppress phonon 
thermal conductance $k_{ph}$~\cite{ztstrain1,ztstrain2,dop1,dop2} that can enhance the value of $ZT$. In addition to these previous 
works, here, we explore another viable route for decreasing $k_{ph}$ based on quasiperiodic localization.

At this stage, it is important to highlight the distinction between `confinement-induced potential modulation' used in 
earlier studies and `explicitly imposed onsite-potential modulation' in our chosen model to make the present study more impactful.  Confinement-induced potential modulation arises from the structural variations that modify the effective potential landscape experienced 
by the charge carriers, whereas in the present model the onsite potential modulation is introduced directly through a spatially varying 
onsite energy. Despite these differences, both can be exploited to modify the transmission spectrum and thereby enhance the thermoelectric performance. The general concept of transmission engineering through deterministic modulation of coherent wave propagation in 
width-modulated nano-waveguides has already been an established idea, but here, we have not used a transmission-engineering strategy. 
Rather, in this work, we propose an idea of controlling electronic and phonon transport by imposing a {\em weakly varying} potential, 
which is a special kind of correlated disorder that produces an asymmetric transmission profile and mobility edges, which can significantly increase the thermoelectric figure of merit ($ZT$).

\section{Experimental Feasibility of Weakly Varying AAH Systems}

It is worth noting that, with sophisticated nanotechnologies, it is possible to design physical systems in the presence of both simple and complex potentials. For AAH-type potentials, experimentalists mostly use optical and ultracold-atom setups\cite{60,61,62}. 
Primarily, two counterpropagating laser beams of different frequencies are used to create the trapping potentials that confine a cloud 
of fermionic atoms. Selectively adjusting the relative frequency of the laser beams, different potential profiles are generated. An 
alternative prescription can also be considered. An array of finger-like gates is formed, and above it, a one-dimensional nanowire is 
placed. With the help of a digital-to-analog converter, voltages in these finger gates can be monitored independently, resulting in a 
desired potential profile. Considering all these possibilities, we strongly believe that our proposed quantum system can be fabricated 
in suitable laboratories. The thermal imbalance $\Delta T$ across the sample can be easily established by means of a micro-heater. 

\section{Closing Remarks and Outlook}

In this study, we propose a method to get large values of $ZT$ at the nanoscopic level, considering a one-dimensional tight-binding chain with weakly varying site potentials connected with two heat reservoirs. Using the Landauer prescription in a tight-binding framework based on NEGF theory, we examine every thermoelectric quantity. Before analyzing the thermoelectric quantities, electronic transmission profiles are discussed elaborately. Our thorough numerical analysis confirms that the conclusions reported here are consistent for a wide parameter range. 
All the characteristic features obtained are summarized below.
\vspace{1pt}\\
\textbullet\ For $\nu=0\,$, i.e., in case of a perfect lattice, all conducting channels are equally accessible. So, a uniform 
electronic transmission spectrum is obtained. An irregular electronic transmission spectrum for $\nu=0.55\,$ indicates that some states 
become localized, due to imposed correlated disorder into the system.\\
\textbullet\ In the presence of correlated disorder, highly non-uniform electronic transmission profile results in large values of Seebeck coefficient, resulting in significantly large $ZT$, although $G$ and $k_e$ become lower relative to the perfect case.\\
\textbullet\ High values of $ZT$ are obtained at those Fermi energies, where $k_e$ is low and $S$ is large.\\
\textbullet\ We discuss the contribution of phonon to thermoelectric efficiency. We calculate phonon transmission probability {\Large$\tau_{\scriptscriptstyle ph}$} using the NEGF technique. Phonon transmission spectra for both perfect and disordered lattices are shown, considering two different sets of systems. Also, we show that disorder reduces $k_{ph}$. Finally, $ZT$ is calculated taking the contributions of both electronic and phonon thermal conductances, and maximum values of $ZT$ exceeding $2$ are obtained for both systems.\\ 
\textbullet\ High values of $ZT$ are found for different system sizes. $ZT$ shows a non-monotonic behavior with temperature, reaching a maximum value at some particular $T$, and then decreases at higher temperatures due to thermal broadening. Favorable thermoelectric response is obtained for different configurations of the disorder and also at different disorder strengths.\\
\textbullet\ We further critically check the impact of coupling asymmetry on $ZT$ and favorable values are obtained. 
$ZT$ is varied simultaneously with $W$ and $\nu$, and large values of $ZT$ are obtained for multiple configurations of the potential, 
which shows the reliability of the system as a efficient thermoelectric material.\\
\textbullet\ Finally, a comparative analysis is done between the $ZT$ values obtained in our system i.e., weakly varying AAH and 
the conventional AAH systems and relatively lower values of $ZT$ are obtained in case of the later one.

The novelty of the work lies in using a weakly varying disorder to tune electron and phonon transport properties for 
enhancing the thermoelectric response rather than engineering thermoelectric transport through deterministic modulation of coherent 
wave propagation. Here, we investigate how the spatial correlations and localization characteristics of this weakly varying potential 
can be explored to obtain favorable thermoelectric responses at the nanoscale regime. We would like to acknowledge that although 
transmission engineering through geometrically modulated thermoelectric metamaterials and nano-waveguides, has already been
established~\cite{nw1,nw2}, here, we propose the implementation of weakly varying AAH potential as an alternative approach for 
obtaining favorable thermoelectric response.

Though the present work involves a simple 1D model to describe the thermoelectric transport characteristics, one can 
extend this study by choosing more realistic and complicated systems. All the underlying physics are expected to remain the same.

The heat baths considered here are modeled as a semi-infinite 1D harmonic chain with fixed atomic masses and spring 
constants with nearest-neighbor harmonic interaction. The central system is coupled to these baths to study the transport characteristics 
and the effects of these heat baths are incorporated through Green's function method. Instead of our chosen idealized 1D baths, we also 
could have considered more realistic and complicated heat baths, but the essential physical pictures are expected to remain unchanged.

Before an end, we would like to point out that, apart from efficient thermoelectric energy conversion, systems with weakly varying 
potentials can be useful to have controlled phase transitions, light confinement, quantum mechanical transport phenomena, different 
topological effects, and to name a few. We will explore these issues in our future work.

\end{document}